%% file: main_arXiv.tex
\documentclass[aps,prapplied,reprint,superscriptaddress]{revtex4-2}

\usepackage{graphicx}
\usepackage{xcolor}
\usepackage{cmap}
\usepackage[T1]{fontenc}
\usepackage{lmodern}
\usepackage{amsmath, amssymb}
\usepackage{multirow}
\definecolor{apslinkblue}{HTML}{2E308C}
\usepackage[colorlinks=true,linkcolor=apslinkblue,citecolor=apslinkblue,urlcolor=apslinkblue]{hyperref}

\begin{document}

\title{Hybrid dynamical decoupling and coherent driving for high-fidelity nuclear-spin control in diamond}

\author{Jiwon Jeon}
\affiliation{Center for Quantum Technology, Korea Institute of Science and Technology, Seoul 02792, Republic of Korea}

\author{Donghun Jung}
\affiliation{Center for Quantum Technology, Korea Institute of Science and Technology, Seoul 02792, Republic of Korea}

\author{Eunsang Lee}
\affiliation{Center for Quantum Technology, Korea Institute of Science and Technology, Seoul 02792, Republic of Korea}

\author{Junghyun Lee}
\affiliation{Center for Quantum Technology, Korea Institute of Science and Technology, Seoul 02792, Republic of Korea}

\date{August 20, 2026}

\begin{abstract}
Nitrogen-vacancy (NV) centers in diamond provide room-temperature electron--nuclear spin registers for quantum sensing and quantum information processing, with surrounding $^{13}$C nuclear spins serving as long-lived quantum memories. However, coherent control of large nuclear-spin registers is limited by finite electron-spin coherence and spectral addressability. Existing approaches follow two complementary strategies: dynamical-decoupling (DD) gates exploit filter-function resonances to realize selective conditional evolution but permit only discrete rotation angles, whereas dynamical-decoupling radio-frequency (DDrf) control restores continuous tunability at the cost of stringent hyperfine-geometry and RF-power requirements. Here, we introduce hybrid dynamical-decoupling and radio-frequency (H-DDrf) control, which preserves the DD-induced conditional evolution and employs a geometrically phase-matched RF drive to complete the target operation. This approach reduces both RF power and gate duration while maintaining high-fidelity control, thereby expanding the accessible $^{13}$C nuclear-spin register for room-temperature NV-based quantum memories and quantum processors.
\end{abstract}

\maketitle

\textit{Introduction.---}Hybrid defect-spin qubits in solid-state materials, formed from coupled electron and nuclear spin networks, provide a promising platform for quantum sensing and quantum information processing~\cite{Doherty2013,Maze2008,Dutt2007,Taminiau2014}. Recent advances in hybrid spin systems have enabled enhanced quantum sensing and long-lived quantum memories through coherent control of coupled electron and nuclear spins~\cite{Maurer2012,Taminiau2012,Shim2013,Waldherr2014,Cramer2016,Bradley2019}. To fully exploit these quantum resources, robust and universal control of the hybrid spin register is essential, particularly at room temperature. In this regime, controllability is often limited by spectral addressability and electron-spin coherence during protected gate sequences~\cite{Hanson2006,Robledo2011,Ryan2010,Bradley2019}. 

\begin{figure}[!t]
\includegraphics[width=\columnwidth]{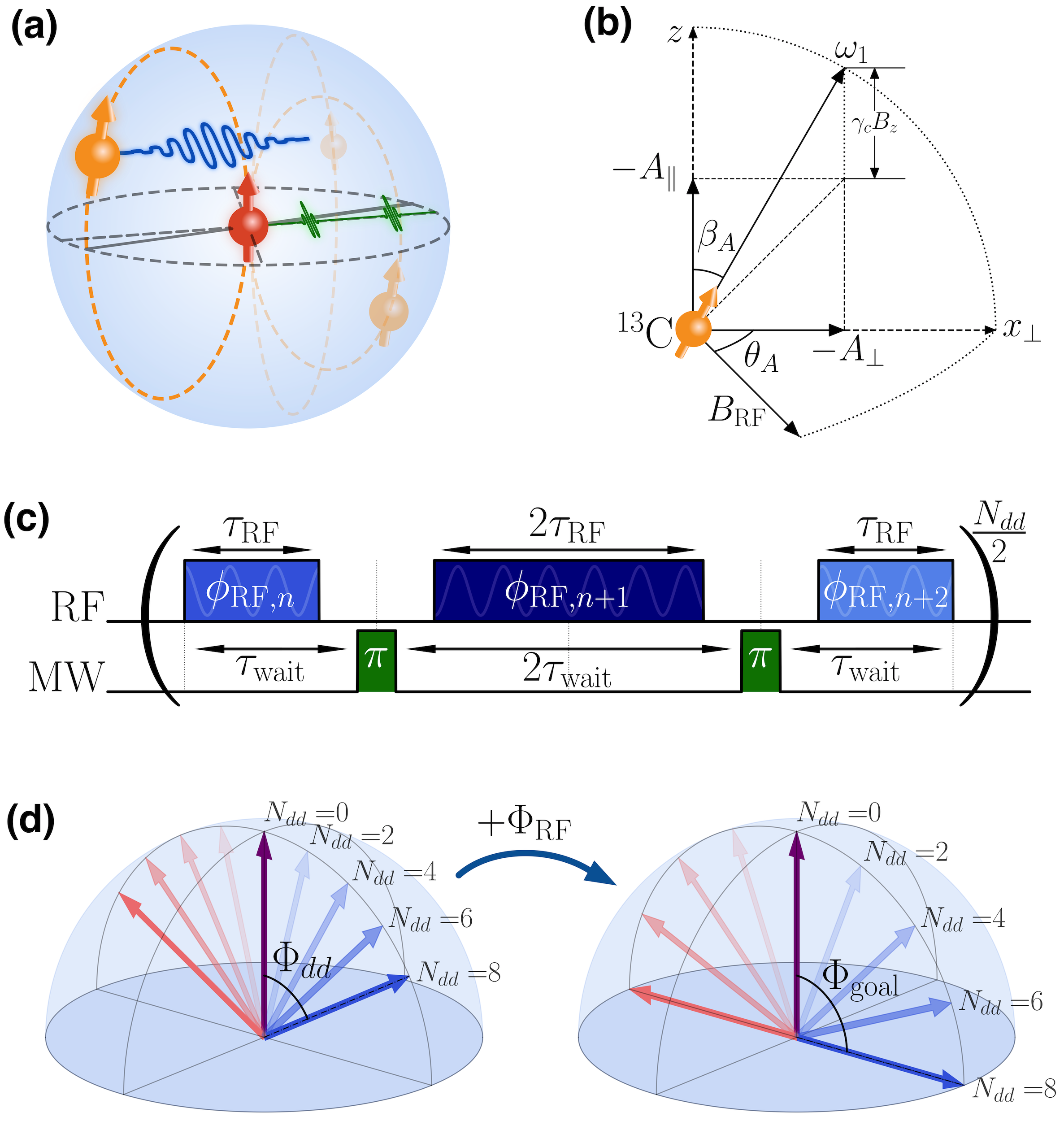}
\caption{\label{fig:overview} Overview of the hybrid dynamical-decoupling radio-frequency (H-DDrf) protocol. (a) Physical implementation of an NV electron spin coupled to a target $^{13}$C nuclear spin, where microwave (MW, green) and radio-frequency (RF, blue) pulses control the electron and nuclear spins, respectively. (b) Hyperfine geometry of the $^{13}\mathrm{C}$ nuclear spin in the $m_s=-1$ manifold, defining the hyperfine azimuthal angle $\theta_A$ and the quantization-axis misalignment $\beta_A$. (c) Pulse-train unit cell showing the RF phase updates $\phi_{\mathrm{RF},n}$, RF duration $\tau_\mathrm{RF}$, and inter-$\pi$-pulse waiting time $\tau_\mathrm{wait}$. (d) Hybridized gate construction on the Bloch hemispheres. The DD-induced rotation $\Phi_{dd}$ $(N_{dd}=8)$ is supplemented by an RF rotation $\Phi_\mathrm{RF}$ to achieve the target rotation $\Phi_\mathrm{goal}$.}
\end{figure}

Nuclear-spin control conditional on the electron-spin state is commonly constrained by precision, addressability, and the size of the accessible nuclear-spin register~\cite{Taminiau2012,Taminiau2014,Abobeih2019,Bradley2019}. The design of robust gate sets critically depends on the geometry and strength of the nitrogen vacancy (NV) center's electron--nuclear hyperfine interaction~\cite{Taminiau2012,Bradley2019,Dong2020}. Existing approaches include dynamical decoupling (DD)-based gates~\cite{Taminiau2012,Taminiau2014,Dong2020}, dynamical-decoupling radio-frequency (DDrf) control~\cite{Bradley2019,VanOmmen2025,Beukers2025}, and direct radio-frequency (RF) driving~\cite{Childress2006,Dutt2007,Shim2013,Robledo2011}. Originally introduced to protect spin coherence~\cite{Carr1954,Meiboom1958}, DD has become a central tool for extending the NV center's electron spin coherence time~\cite{Ryan2010,deLange2010,Naydenov2011,ShimDD2012,BarGill2013}. DD-based controlled two-qubit gates, typically implemented using Carr-Purcell-Meiboom-Gill (CPMG) sequences~\cite{Carr1954,Meiboom1958,Gullion1990,Souza2011}, are highly versatile but constrain the rotation angle to discrete values determined by the even-integer number of DD $\pi$-pulses~\cite{Taminiau2012,Taminiau2014,Dong2020}. Introducing detuning in the inter-pulse spacing can alleviate this limitation by enabling continuous control of the rotation angle, but it also induces additional undesired phase rotations that complicate gate design~\cite{Casanova2015,Wang2016,Casanova2016,Dong2020}. The DDrf approach has demonstrated high gate fidelities and offers enhanced nuclear-spin selectivity~\cite{Bradley2019,VanOmmen2025,Beukers2025}. However, existing DDrf analyses and implementations have largely focused on systems with weak transverse hyperfine couplings or hyperfine tensors aligned with the quantization axis, with gate performance constrained by RF-pulse bandwidth, off resonance in driving, and gate-duration trade-offs~\cite{VanOmmen2025,Beukers2025}. Meanwhile, direct RF control enables fast nuclear-spin gates but is generally limited to a few strongly coupled nuclear spins near the electron spin~\cite{Childress2006,Dutt2007,Shim2013,Robledo2011}. Extending this approach to weakly coupled nuclear spins is generally impractical, as the required strong RF driving for fast control leads to substantial off-resonant cross-talks~\cite{VanOmmen2025,Beukers2025}. 

Motivated by these limitations, we introduce a hybrid DDrf (H-DDrf) gate protocol in which the DD-induced conditional rotation (CR) is retained as a useful coherent offset rather than suppressed. In H-DDrf, the DD sequence selectively addresses the target nuclear spin and generates a discrete conditional operation, while an interleaved RF drive provides the remaining rotation with a phase matched to the DD-induced rotation axis. This approach shifts the role of RF control from replacing DD-induced dynamics to completing them, enabling continuous rotation-angle tunability while preserving DD-based selectivity. In addition, H-DDrf explicitly compensates hyperfine-geometry-dependent effects, enabling nuclear-spin control over a broader range of hyperfine tensors.

\textit{Hybrid DDrf protocol.---}We first present the theoretical framework underlying the proposed H-DDrf protocol and derive the conditions required for its implementation. A microwave (MW) drive addresses the NV electron-spin transition, while periodic MW $\pi$-pulses within a DD sequence enable conditional control of the nuclear spin through the anisotropic hyperfine interaction. This condition is achieved by choosing an appropriate inter-$\pi$-pulse train period $T_\mathrm{train} = 2\tau_\mathrm{wait} + \tau_\mathrm{MW}$~\cite{Taminiau2012,Bradley2019,Dong2020}, where $\tau_\mathrm{MW}$ is the finite MW $\pi$-pulse duration. Consequently, the DD-induced CR angle $\Phi_{dd}$ is quantized because the rotation accumulates discretely with the number of DD $\pi$-pulses $N_{dd}$. In contrast, conventional DDrf protocols choose $T_\mathrm{train}$ to suppress $\Phi_{dd}$, allowing the nuclear spin dynamics to be driven predominantly by the RF field~\cite{Bradley2019,VanOmmen2025,Beukers2025}. Here, instead, $T_\mathrm{train}$ is chosen to preserve the DD-induced conditional evolution, while a geometrically phase-matched RF drive supplies the remaining rotation. The resulting RF-assisted rotation angle satisfies
\begin{equation}\label{eq:rf-angle}
    \Phi_\mathrm{RF} = \Phi_\mathrm{goal} - \Phi_{dd},
\end{equation}
where the target conditional-rotation angle is typically chosen as $\Phi_\mathrm{goal} = (2n-1)\pi/2$ with $n\in\mathbb{Z}^+$.

Using the $m_s=0$ and $-1$ NV states as an effective two-level system, we describe the electron spin in the MW rotating frame and the nuclear spin in the laboratory frame. The free Hamiltonian of an NV center hyperfine coupled to, for example, a single $^{13}\mathrm{C}$ nuclear spin is~\cite{Doherty2013,Bradley2019},
\begin{equation}
    \hat{H}_0 = \gamma_c B_z \hat{I}_z + A_\parallel \hat{S}_z \hat{I}_z + A_\perp\hat{S}_z(\hat{I}_x\cos\theta_A + \hat{I}_y\sin\theta_A),
\end{equation}
where $\gamma_c/2\pi = 1.0705\ \mathrm{kHz/G}$ is the $^{13}\mathrm{C}$ gyromagnetic ratio, $B_z$ is the static magnetic field aligned with the NV axis, and $A_\parallel$ and $A_\perp$ are the longitudinal and transverse hyperfine coupling strengths, respectively. The corresponding hyperfine geometry and nuclear quantization axes are illustrated in Fig.~\ref{fig:overview}(b). The azimuthal angle $\theta_A$ specifies the direction of the transverse hyperfine field, with the $x$-axis defined by the applied RF magnetic field, $B_\mathrm{RF}$. The nuclear Larmor frequencies in the $m_s = 0$ and $-1$ manifolds are $\omega_0 = \gamma_c B_z$ and $\omega_1 = \sqrt{(\omega_0 - A_\parallel)^2 + A_\perp^2}$, respectively, with mean frequency $\bar{\omega} = (\omega_0 + \omega_1)/2$. We further define the hyperfine-induced detuning as $\delta_A = \omega_1 - \omega_0$ and the quantization-axis tilt in the $m_s=-1$ manifold as $\beta_A = \sin^{-1}(A_\perp/\omega_1)$.

Previous treatments typically assume weak transverse hyperfine coupling, $A_\perp$, or neglect the effects of finite transverse hyperfine components to simplify the control protocol or theoretical analysis~\cite{Bradley2019,VanOmmen2025,Finsterhoelzl2025}. Here, we explicitly retain the finite quantization-axis misalignment while operating at moderate magnetic fields. A non-negligible $\beta_A$ produces two geometric effects in the $m_s = -1$ manifold: it reduces the effective RF Rabi frequency, thereby increasing the RF amplitude required for a target operation, and shifts the phase of the RF drive experienced by the nuclear spin. These effects are captured by the complex geometric factor,
\begin{equation}\label{eq:geom-factor}
    \mathcal{C}_\phi = \sin^2\theta_A + \cos\beta_A\cos^2\theta_A + i\sin\theta_A\cos\theta_A(\cos\beta_A - 1),
\end{equation}
whose magnitude, $R_A = |\mathcal{C}_\phi|$, defines the RF pulse-dilation factor, while its phase, $\delta\phi = \mathrm{arg}(\mathcal{C}_\phi)$, gives the geometric phase shift in the $m_s = -1$ manifold. In practice, $R_A$ determines the RF pulse dilation required to achieve the target rotation angle, whereas $\delta\phi$ specifies the initial phase correction needed to align the RF- and DD-induced rotation axes.

To implement the H-DDrf protocol, the pulse-train period is chosen as $T_\mathrm{train} = \pi(2k_{dd} - 1)/\bar{\omega}$, with $k_{dd}\in\mathbb{Z}^+$, satisfying the DD-induced CR resonance condition~\cite{Taminiau2012,Beukers2025}. For a given $N_{dd}$, the RF Rabi frequency is then set by $\Omega_\mathrm{RF} = \Phi_\mathrm{RF}/(N_{dd}\tau_\mathrm{RF})$, where $\tau_\mathrm{RF}$ is the RF pulse duration and satisfies $2\tau_\mathrm{RF}\leq T_\mathrm{train}$. In the $m_s = 0$ manifold, the nuclear spin simultaneously undergoes an off-resonant Rabi oscillation with detuning $\delta_A$, which reduces the gate fidelity. To suppress this unwanted evolution, the RF pulse duration is chosen such that the detuned oscillation completes an integer number of cycles, and choosing the fastest solution gives:
\begin{equation}\label{eq:rf-duration}
    \tau_\mathrm{RF} = \frac{\sqrt{4\pi^2N_{dd}^2R_A^2 - \Phi_\mathrm{RF}^2}}{|\delta_A|N_{dd}R_A}.
\end{equation}
This constraint sets the minimum pulse-train index, $k_{dd,\mathrm{min}} = \lceil 2\bar{\omega}/|\delta_A| + 1/2\rceil$, required to satisfy $2\tau_\mathrm{RF}\leq T_\mathrm{train}$. With the pulse duration fixed, the RF phase must be updated to compensate the relative phase accumulated from the different nuclear Larmor frequencies in the two electron-spin manifolds. Following the arithmetic phase-update protocol of conventional DDrf control~\cite{Bradley2019,VanOmmen2025,Beukers2025}, we use
\begin{equation}\label{eq:phase-update}
      \phi_{\mathrm{RF},n_{dd}} = n_{dd}\left(\pi - \frac{\delta_A T_\mathrm{train}}2\right) + \phi_0,
\end{equation}
where $\phi_0 = \theta_A - \delta\phi$ is the initial phase that aligns the RF and DD-induced rotation axes. The alternating $\pi$-phase shift ensures that the RF drive acts as a conditional counter-rotation in successive DD intervals. Together, these relations specify the RF rotation angle, pulse duration, and phase sequence used in the simulations below.

\begin{figure}[!t]
\includegraphics[width=\columnwidth]{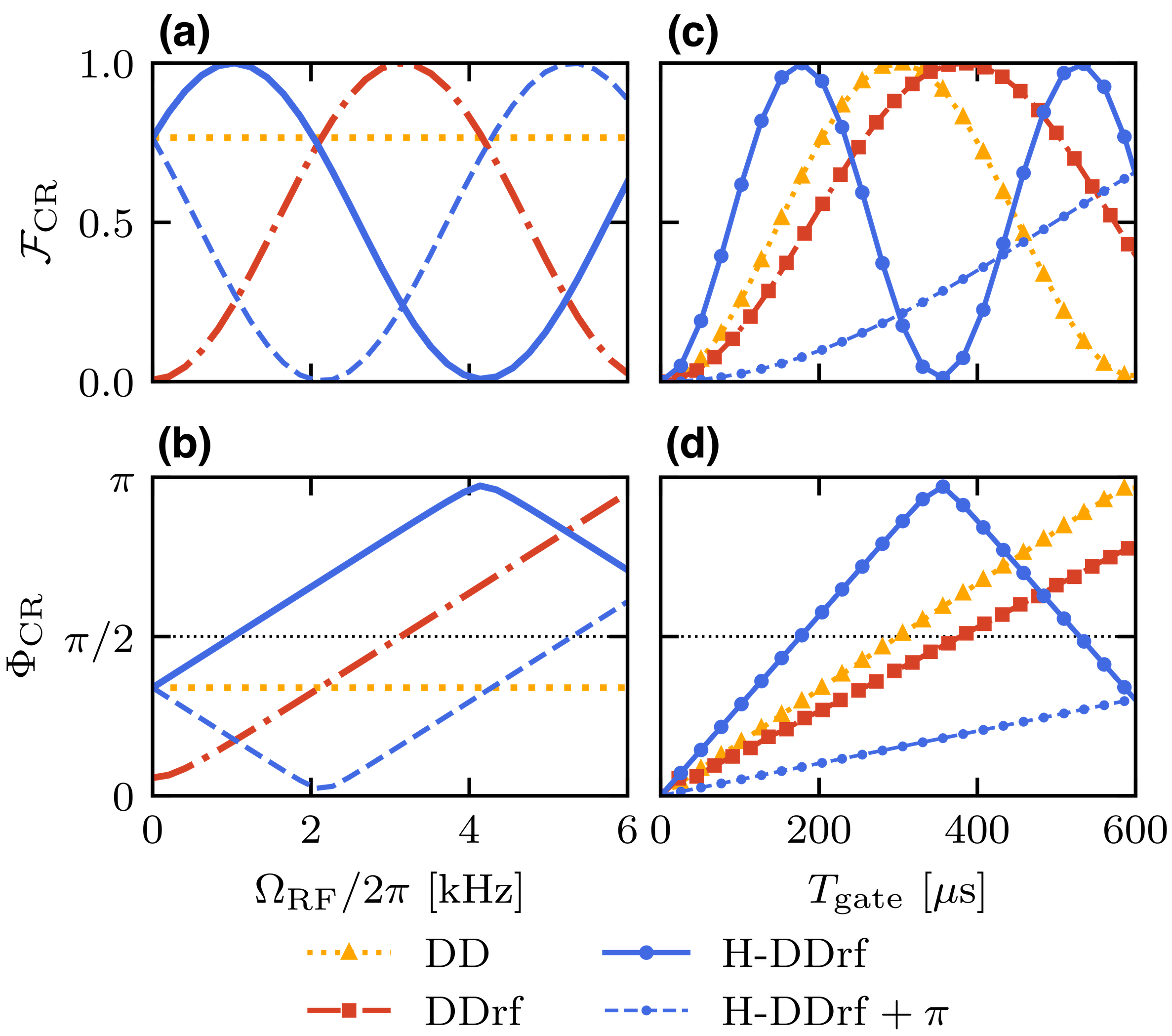}
\caption{\label{fig:temp2} Conditional-rotation (CR) fidelity $\mathcal{F}_\mathrm{CR}$ and rotation angle $\Phi_\mathrm{CR}$ for DD, DDrf, and H-DDrf control of a target spin with $(A_\parallel,A_\perp)/2\pi=(-200,50)\ \mathrm{kHz}$ and $\theta_A=0$. (a),(b) $\mathcal{F}_\mathrm{CR}$ and $\Phi_\mathrm{CR}$ as functions of RF Rabi frequency, $\Omega_\mathrm{RF}$, for fixed $N_{dd}=16$. (c),(d) $\mathcal{F}_\mathrm{CR}$ and $\Phi_\mathrm{CR}$ as functions of gate duration, $T_\mathrm{gate}=N_{dd}T_\mathrm{train}$, for even $N_{dd}$ at fixed $\Omega_\mathrm{RF}/2\pi = 1.50\ \mathrm{kHz}$. Hybrid+$\pi$ denotes H-DDrf with the initial RF phase shifted by $\pi$.}
\end{figure}

\textit{Control performance.---}We evaluate the gate performance for a static magnetic field of $B_z=440.1\ \mathrm{G}$ applied along the NV axis, using numerical propagators with finite MW $\pi$-pulses, $T_\mathrm{train}=2\tau_\mathrm{wait}+\tau_\mathrm{MW}$, and finite RF pulse windows. Throughout this work, the conditional-operation contrast $\mathcal{F}_\mathrm{CR} = 1 - |\mathrm{Tr}(\hat{\mathcal{U}}_{-1}^{\dagger}\hat{\mathcal{U}}_0)|^2/4$ is used as the figure of merit for the CR gate fidelity, where $\hat{U}\simeq\hat{\mathcal{U}}_0\oplus\hat{\mathcal{U}}_{-1}$ denotes the block-diagonal nuclear-spin evolution in the $m_s=0$ and $m_s=-1$ manifolds~\cite{Nielsen2002,Pedersen2007}. The corresponding CR angle, $\Phi_\mathrm{CR}$, is extracted from the nuclear-spin rotations generated by $\hat{\mathcal{U}}_0$ and $\hat{\mathcal{U}}_{-1}$. Details of the numerical propagation, RF and MW pulse parameters, and DDrf parameter-selection procedure are provided in the Supplemental Material.

Fig.~\ref{fig:temp2} illustrates the controllability and tunability of DD-only, DDrf, and H-DDrf control for a representative target nuclear spin. In the RF-power and gate-duration scans, the RF pulse duration is recalculated from Eq.~\ref{eq:rf-duration} for each operating point, while the DDrf comparison uses the pulse parameters specified in the Supplemental Material. At fixed $N_{dd}=16$, DD generates a finite CR angle, which H-DDrf treats as a coherent offset and supplements with an RF-driven rotation according to Eq.~\ref{eq:rf-angle}. By contrast, DDrf suppresses the DD-induced evolution, such that the CR operation is generated entirely by the RF drive~\cite{Bradley2019,VanOmmen2025,Beukers2025}. As shown in Fig.~\ref{fig:temp2}, the power dependence demonstrates that H-DDrf achieves the target CR angle with a reduced RF Rabi frequency, while the opposite-phase branches enable bidirectional tuning around the DD-defined operating point. At fixed RF power, Fig.~\ref{fig:temp2}(c),(d) further shows that the target operation is reached with fewer DD $\pi$-pulses and a shorter gate duration while preserving high CR fidelity.

\begin{figure}[!t]
\includegraphics[width=\columnwidth]{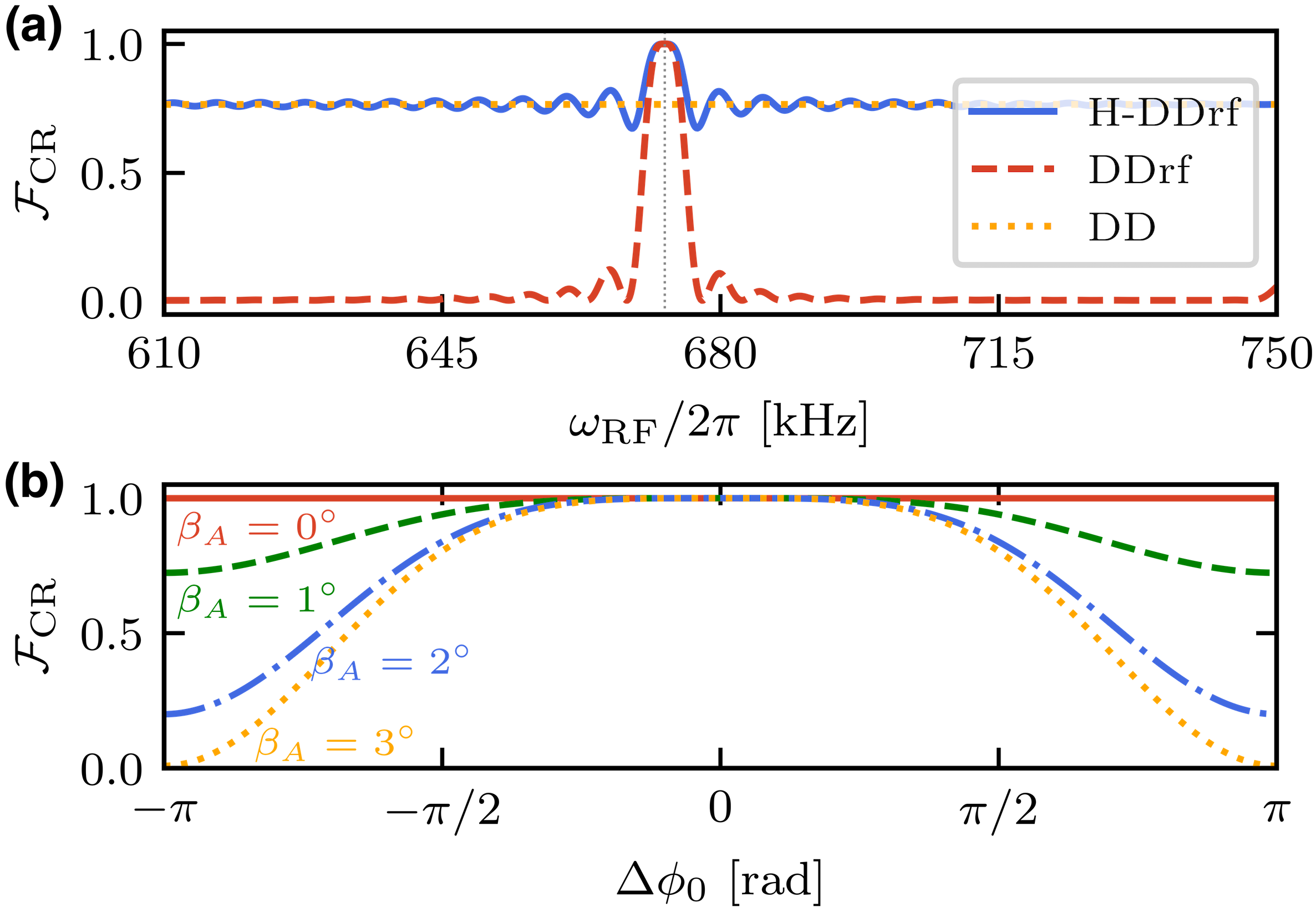}
\caption{\label{fig:selectivity}Frequency selectivity and geometric phase matching of H-DDrf. (a) CR fidelity $\mathcal{F}_\mathrm{CR}$ under H-DDrf (blue), DDrf (dashed red), and DD-only (dotted orange) control for a target $^{13}$C spin with $(A_\parallel,A_\perp)/2\pi=(-200,50)\ \mathrm{kHz}$, $\theta_A=0$, and $N_{dd}=16$, as a function of the RF drive frequency $\omega_\mathrm{RF}$. The DDrf parameters are chosen as in Fig.~\ref{fig:temp2}. (b) $\mathcal{F}_\mathrm{CR}$ under H-DDrf as a function of the initial RF phase for nuclear spins with fixed $\delta_A/2\pi=220\ \mathrm{kHz}$ and varying $\beta_A\in\{0^\circ,1^\circ,2^\circ,3^\circ\}$. The phase offset, $\Delta\phi$, is defined relative to the geometrically matched initial phase $\phi_0$, given by Eq.~\eqref{eq:phase-update}.}
\end{figure}

We next examine whether the RF-assisted contribution preserves selectivity to the target nuclear spin. In Fig.~\ref{fig:selectivity}(a), the same target spin and $N_{dd}=16$ are used as in Fig.~\ref{fig:temp2}, while only the applied RF frequency $\omega_\mathrm{RF}$ is swept. H-DDrf enhances the CR near the target transition frequency $\omega_1$, while the response remains close to the DD-induced offset away from resonance. Thus, the DD sequence establishes the frequency-selective CR baseline~\cite{Taminiau2012,Dong2020}, and the RF drive provides a resonant correction rather than replacing the DD-selected operation. This distinction is crucial in multi-spin environments, where H-DDrf preserves DD-based selectivity while adding continuous RF tunability. 

The RF-assisted rotation must also be phase matched to the DD-induced evolution. The DD rotation axis is determined by the hyperfine azimuthal angle $\theta_A$, whereas the RF drive in the $m_s=-1$ manifold requires an additional phase shift $\delta\phi$ from Eq.~\ref{eq:geom-factor}. Consequently, the phase condition $\phi_0=\theta_A-\delta\phi$ in Eq.~\ref{eq:phase-update} aligns the RF and DD rotation axes. Fig.~\ref{fig:selectivity}(b) tests this condition by scanning the applied initial RF phase around $\phi_0$ for spins with fixed $\delta_A/2\pi=220\ \mathrm{kHz}$ and varying $\beta_A=0^\circ$--$3^\circ$. The correction becomes significantly important with increasing $\beta_A$: for small $\beta_A$, the gate fidelity is nearly insensitive to the RF phase, whereas larger quantization-axis misalignment substantially suppresses the CR fidelity unless the phase-matching condition is satisfied. This demonstrates that the phase correction is a geometric requirement arising from the hyperfine interaction, rather than an empirical fitting parameter. 

\begin{figure}[!t]
\includegraphics[width=\columnwidth]{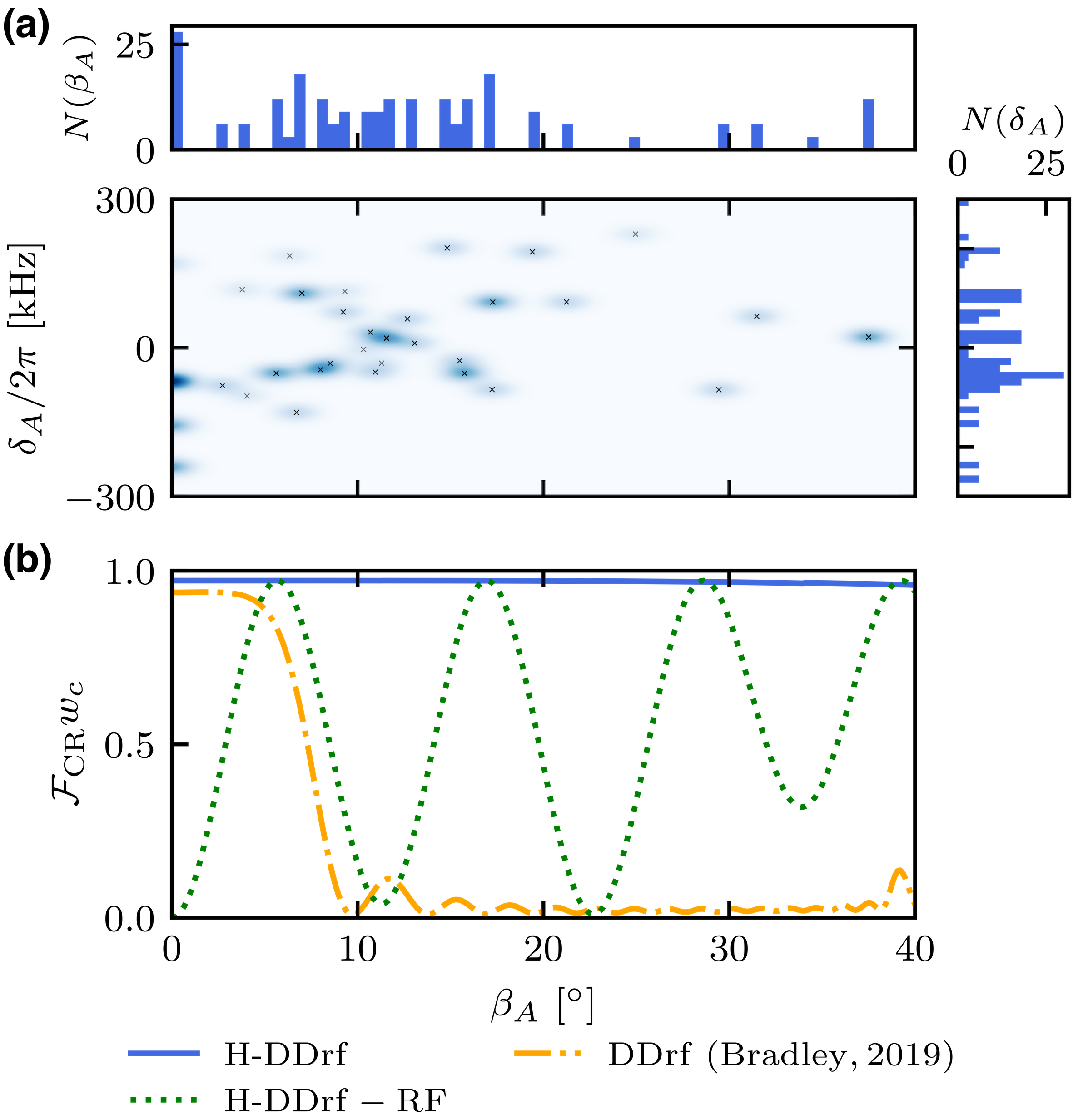}
\caption{\label{fig:scalability}
Hyperfine-parameter distribution and coherence-weighted operation fidelity. (a) Distribution of possible hyperfine parameters in the $\delta_A$--$\beta_A$ phase space, obtained from a diamond-lattice survey assuming 100\% $^{13}\mathrm{C}$ abundance. For carbon sites 5--7 \AA{} from the NV center, the parameters are estimated using the secular electron-nuclear magnetic dipole-dipole interaction~\cite{Doherty2013,Bradley2019}. (b) Coherence-weighted CR fidelity, $\mathcal{F}_\mathrm{CR}w_c$, for H-DDrf (blue), H-DDrf without RF assistance (green dotted), and DDrf following Ref.~\cite{Bradley2019} (orange dash-dotted), evaluated for nuclear spins with $\delta_A/2\pi=220\ \mathrm{kHz}$ and $\beta_A\in[0^\circ,40^\circ]$. The vertical dotted line marks the onset of fidelity degradation in DDrf. Here, $w_c$ accounts for NV electron-spin coherence loss during the finite DD-protected gate. The coherence model follows standard DD-protected NV coherence envelopes~\cite{deLange2010,Ryan2010,Naydenov2011,ShimDD2012}, with parameters chosen consistently with recent DDrf analysis~\cite{Beukers2025}.}
\end{figure}

\textit{Scalability and outlook.---}H-DDrf broadens the accessible hyperfine-parameter regime by compensating finite transverse hyperfine couplings rather than avoiding them. Fig.~\ref{fig:scalability}(a) is obtained from a diamond-lattice survey of carbon sites 5--7~\AA{} from the NV center, treating each site as a possible $^{13}\mathrm{C}$ register and estimating the hyperfine parameters from the electron-spin-secular magnetic dipole-dipole interaction~\cite{Doherty2013,Bradley2019}. The dipolar hyperfine interaction naturally gives rise to $^{13}\mathrm{C}$ sites with substantial transverse components over the relevant coupling range. These spins are challenging for protocols that assume weak transverse coupling or suppress its effects~\cite{Bradley2019,VanOmmen2025,Finsterhoelzl2025}. In contrast, H-DDrf incorporates the geometric factor $\mathcal{C}_\phi$ [Eq.~\ref{eq:geom-factor}] to align the RF-assisted rotation with the DD-induced conditional evolution, thereby maintaining high-fidelity control across a broader range of hyperfine geometries. Fig.~\ref{fig:scalability}(b) fixes $\delta_A/2\pi=220\ \mathrm{kHz}$ and varies $\beta_A=0^\circ$--$40^\circ$, plotting the coherence-weighted fidelity $\mathcal{F}_\mathrm{CR}w_c$ with $w_c$ accounting for NV coherence loss during the finite gate duration. The H-DDrf response remains high over a broad range of $\beta_A$, whereas DDrf following Ref.~\cite{Bradley2019} exhibits a marked degradation in the large-misalignment regime. Here, DDrf is implemented using the parameter-selection procedure of Ref.~\cite{Bradley2019}, rather than the optimized conditions adopted for the comparison in Figs.~\ref{fig:temp2} and \ref{fig:selectivity}. This contrast illustrates the central mechanism of H-DDrf: the DD sequence provides frequency selectivity and a discrete conditional-rotation baseline, while the phase-matched RF drive coherently completes the target operation without replacing the DD-induced dynamics. 

In conclusion, we establish H-DDrf as a control strategy that exploits DD-induced dynamics and RF driving as complementary resources. By combining the selectivity of DD with geometrically phase-matched RF control, the protocol reduces RF-power and gate-time requirements while maintaining high-fidelity operation across a broader range of controllable $^{13}$C nuclear spins. The reduced RF-power requirement and shorter gate duration of H-DDrf are particularly advantageous for room-temperature NV experiments. Shorter gates reduce exposure to finite electron-spin coherence times~\cite{deLange2010,Ryan2010,Naydenov2011,BarGill2013}, while weaker RF driving mitigates technical limitations associated with strong control fields, including drive-induced heating, amplitude noise, and off-resonant excitation~\cite{VanOmmen2025,Beukers2025,Hu2024}. H-DDrf, however, does not eliminate the intrinsic selectivity constraints shared by DD- and DDrf-based control. DD filter functions exhibit higher-order harmonic resonances allowing nuclear spins with different Larmor frequencies to satisfy resonance conditions~\cite{Taminiau2012,Casanova2015,Dong2020,Loretz2015}, while RF driving can produce additional off-resonant excitation peaks~\cite{Bradley2019,VanOmmen2025}. These effects must be considered when selecting $T_\mathrm{train}$, $\omega_\mathrm{RF}$, and the target nuclear-spin register. At the same time, they suggest a potential route toward multi-qubit gate control, in which the DD- and RF-assisted components selectively address different spins. Additional simulations in the Supplemental Material examine the deterministic parameter-selection procedure, harmonic responses, RF-control robustness, and the limiting case where the DD-induced CR nearly completes the target operation without RF assistance. Exploring such multi-target protocols, however, lies beyond the scope of the present work. 

Achieving high selectivity, fidelity, and scalability in nuclear-spin control under ambient conditions is essential for realizing quantum error correction and error mitigation in advanced quantum technologies. In diamond, nearby nuclear spins have been employed as ancillary and memory qubits for stabilizer measurements, repeated correction, and feedback-protected logical states~\cite{Taminiau2014,Waldherr2014,Cramer2016}, representing key steps toward fault-tolerant defect-spin quantum memories. By expanding the accessible $^{13}$C nuclear-spin register under ambient conditions while reducing RF-power and gate-time requirements, H-DDrf provides additional controllable nuclear-spin resources for such protected multi-qubit protocols.

\begin{acknowledgments}
This work was supported by the National Research Foundation of Korea (Grant No. RS-2024-00442710 and RS-2025-25454922), the Institute for Information \& Communications Technology Planning \& Evaluation (Grant No. RS-2025-25464252 and RS-2025-02219034) and the KIST institutional program (Grant No. 26E001 and 26E0011).
\end{acknowledgments}

\input{main_references.bbl}
\input{supp_after_main.tex}

\begingroup
\phantomsection
\label{sec:supp-references}
\makeatletter
\let\hddrf@label\label
\renewcommand{\label}[1]{%
  \def\hddrf@tempa{#1}%
  \def\hddrf@tempb{LastBibItem}%
  \ifx\hddrf@tempa\hddrf@tempb\else\hddrf@label{#1}\fi}

\input{supp_references.bbl}
\endgroup

\end{document}

%% file: supp_after_main.tex
\clearpage
\onecolumngrid
\hypersetup{pageanchor=false}
\setcounter{page}{1}
\setcounter{section}{0}
\setcounter{subsection}{0}
\setcounter{subsubsection}{0}
\setcounter{equation}{0}
\setcounter{figure}{0}
\setcounter{table}{0}
\renewcommand{\theequation}{S\arabic{equation}}
\renewcommand{\thefigure}{S\arabic{figure}}
\renewcommand{\thetable}{S\arabic{table}}
\renewcommand{\theHequation}{supp.equation.\arabic{equation}}
\renewcommand{\theHfigure}{supp.figure.\arabic{figure}}
\renewcommand{\theHtable}{supp.table.\arabic{table}}
\renewcommand{\theHsection}{supp.section.\arabic{section}}
\renewcommand{\theHsubsection}{supp.subsection.\arabic{section}.\arabic{subsection}}
\renewcommand{\theHsubsubsection}{supp.subsubsection.\arabic{section}.\arabic{subsection}.\arabic{subsubsection}}
\setcounter{tocdepth}{2}

\begin{center}
\parbox{0.88\textwidth}{\centering\large\bfseries\setlength{\baselineskip}{17pt}\selectfont Supplemental Material: Hybrid dynamical decoupling and coherent driving for high-fidelity nuclear-spin control in diamond}\par
\vspace{0.8em}
\renewcommand{\baselinestretch}{1.18}\selectfont
Jiwon Jeon$^{1}$, Donghun Jung$^{1}$, Eunsang Lee$^{1}$, and Junghyun Lee$^{1}$\par
{\it $^{1}$Center for Quantum Technology, Korea Institute of Science and Technology, Seoul 02792, Republic of Korea}\par
\vspace{0.6em}
(Dated: August 20, 2026)
\end{center}

\begin{center}
\renewcommand{\baselinestretch}{1.15}\selectfont
\textbf{CONTENTS}
\end{center}
\begingroup
\newcommand{\tocgap}{\\[0.75em]}
\noindent\hyperref[sec:supp-theory]{I. Theoretical framework}\hfill\pageref{sec:supp-theory}\tocgap
\hspace*{1.5em}\hyperref[sec:supp-frame]{A. Frame convention and system Hamiltonian}\hfill\pageref{sec:supp-frame}\tocgap
\hspace*{1.5em}\hyperref[sec:supp-hyperfine-geometry]{B. Hyperfine geometry and nuclear quantization axes}\hfill\pageref{sec:supp-hyperfine-geometry}\tocgap
\hspace*{1.5em}\hyperref[sec:supp-dd-cr]{C. DD-induced conditional rotation}\hfill\pageref{sec:supp-dd-cr}\tocgap
\hspace*{1.5em}\hyperref[sec:supp-geometric-factor]{D. Geometric factor for RF driving}\hfill\pageref{sec:supp-geometric-factor}\tocgap
\hspace*{1.5em}\hyperref[sec:supp-rf-unitary]{E. RF pulse unitary operator}\hfill\pageref{sec:supp-rf-unitary}\tocgap
\hspace*{1.5em}\hyperref[sec:supp-rf-duration]{F. RF pulse-duration condition}\hfill\pageref{sec:supp-rf-duration}\tocgap
\hyperref[sec:supp-simulation]{II. Simulation and evaluation procedure}\hfill\pageref{sec:supp-simulation}\tocgap
\hspace*{1.5em}\hyperref[sec:supp-propagator]{A. Total DDrf sequence propagator}\hfill\pageref{sec:supp-propagator}\tocgap
\hspace*{1.5em}\hyperref[sec:supp-cr-fidelity]{B. Conditional-operation contrast and CR-angle extraction}\hfill\pageref{sec:supp-cr-fidelity}\tocgap
\hspace*{1.5em}\hyperref[sec:supp-coherence-weight]{C. Coherence-weight model}\hfill\pageref{sec:supp-coherence-weight}\tocgap
\hspace*{1.5em}\hyperref[sec:supp-rf-selection]{D. RF-assisted rotation magnitude and sign selection}\hfill\pageref{sec:supp-rf-selection}\tocgap
\hspace*{1.5em}\hyperref[sec:supp-main-fig-parameters]{E. Detailed parameter settings for main-text figures}\hfill\pageref{sec:supp-main-fig-parameters}\tocgap
\hspace*{1.5em}\hyperref[sec:supp-bradley19-strategy]{F. Original-DDrf parameter-selection landscape}\hfill\pageref{sec:supp-bradley19-strategy}\tocgap
\hyperref[sec:supp-additional-results]{III. Additional simulation results}\hfill\pageref{sec:supp-additional-results}\tocgap
\hspace*{1.5em}\hyperref[sec:supp-time-domain]{A. Time-domain evolution simulations}\hfill\pageref{sec:supp-time-domain}\tocgap
\hspace*{1.5em}\hyperref[sec:supp-gate-level]{B. Gate-level unitary simulations}\hfill\pageref{sec:supp-gate-level}\tocgap
\hspace*{3em}\hyperref[sec:supp-hddrf-performance-section]{1. Broad hyperfine-geometry applicability of H-DDrf}\hfill\pageref{sec:supp-hddrf-performance-section}\tocgap
\hspace*{3em}\hyperref[sec:supp-no-rf-region]{2. RF-unnecessary regions from DD-only completion}\hfill\pageref{sec:supp-no-rf-region}\tocgap
\hspace*{3em}\hyperref[sec:supp-selective-harmonic]{3. Selective enhancement and harmonic responses}\hfill\pageref{sec:supp-selective-harmonic}\tocgap
\hspace*{1.5em}\hyperref[sec:supp-robustness]{C. Representative robustness checks}\hfill\pageref{sec:supp-robustness}\tocgap
\hyperref[sec:supp-diamond-survey]{IV. Diamond-lattice hyperfine-geometry survey}\hfill\pageref{sec:supp-diamond-survey}\tocgap
\hyperref[sec:supp-references]{References}\hfill\pageref{sec:supp-references}
\endgroup
\clearpage

\section{Theoretical framework}\label{sec:supp-theory}

\subsection{Frame convention and system Hamiltonian}\label{sec:supp-frame}
The laboratory-frame Hamiltonian of a single NV center in the triplet ground state coupled to a single $^{13}\mathrm{C}$ nuclear spin under a static magnetic field $B_z$ along the NV symmetry axis is described as~\cite{supp:Childress2006,supp:Dutt2007,supp:Doherty2013}
\begin{equation}\label{eq:supp-lab-hamiltonian}
    \hat{H}_{\mathrm{lab}} = \Delta_{\mathrm{ZFS}}\hat{S}_z^2 + \gamma_eB_z\hat{S}_z + \gamma_cB_z\hat{I}_z + A_\parallel\hat{S}_z\hat{I}_z + A_\perp\hat{S}_z(\hat{I}_x\cos\theta_A+\hat{I}_y\sin\theta_A).
\end{equation}
We set $\hbar=1$ and use angular-frequency units. The quantity $\Delta_{\mathrm{ZFS}}/2\pi\simeq2.87~\mathrm{GHz}$ is the zero-field splitting of the NV triplet ground state, $\gamma_e/2\pi\simeq2.803~\mathrm{MHz/G}$ is the electron gyromagnetic ratio, and $\gamma_c/2\pi\simeq1.0705~\mathrm{kHz/G}$ is the $^{13}\mathrm{C}$ nuclear gyromagnetic ratio. The hyperfine parameters $A_\parallel$ and $A_\perp$ are the longitudinal and transverse components of the coupling, respectively, and $\theta_A$ is the azimuthal direction of the transverse hyperfine component.

The microwave (MW) drive is tuned to the $m_s=0\leftrightarrow -1$ transition with frequency $\omega_{\mathrm{MW}}=\Delta_{\mathrm{ZFS}}-\gamma_eB_z$~\cite{supp:Doherty2013,supp:Taminiau2012}. The frame transformation is applied only to the electron spin, while the nuclear spin is kept in the laboratory frame. The static Hamiltonian in this electron-spin rotating frame is
\begin{equation*}
    \hat{H}_{e\text{-}\mathrm{rot}} = \Delta_{\mathrm{ZFS}}(\hat{S}_z^2+\hat{S}_z) + \gamma_cB_z\hat{I}_z + A_\parallel\hat{S}_z\hat{I}_z + A_\perp\hat{S}_z(\hat{I}_x\cos\theta_A+\hat{I}_y\sin\theta_A).
\end{equation*}
In the same frame, the resonant MW control Hamiltonian becomes $\hat{H}_{\mathrm{MW}}=\Omega_{\mathrm{MW}}\hat{S}_x/\sqrt{2}$.

We use the two-level electron-spin subspace $\{|0\rangle,|\!-\!1\rangle\}$, as commonly used for NV--nuclear-spin control~\cite{supp:Taminiau2012,supp:Taminiau2014,supp:Bradley2019}. In this subspace,
\begin{equation*}
\hat{S}_z \to \frac12\left(\hat\sigma_z - \hat{\mathbb{I}}_2\right),
\qquad
\hat{S}_x\rightarrow\frac{\hat{\sigma}_x}{\sqrt{2}},
\end{equation*}
where $\hat{\sigma}_{x,z}$ are the Pauli $SU(2)$ operators and $\hat{\mathbb{I}}_2$ is the identity operator in the reduced electron-spin subspace. This gives $\hat{S}_z^2+\hat{S}_z=0$, so the system Hamiltonian in the electron-spin rotating frame while keeping the nuclear spin in the laboratory frame is
\begin{equation*}
    \hat{H}_0 = \gamma_c B_z \hat{I}_z + A_\parallel \hat{S}_z \hat{I}_z + A_\perp \hat{S}_z(\hat{I}_x\cos\theta_A+\hat{I}_y\sin\theta_A),
\end{equation*}
which is shown as Eq.~(2) in the main text.

\subsection{Hyperfine geometry and nuclear quantization axes}\label{sec:supp-hyperfine-geometry}
The system Hamiltonian can equivalently be written in the electron-qubit block-diagonal form $\hat{H}_0=\hat{\mathcal{H}}_{0}^{(0)} \oplus \hat{\mathcal{H}}_{0}^{(-1)}$, following the conditional nuclear-spin Hamiltonian used in NV spin-register control~\cite{supp:Taminiau2012,supp:Bradley2019,supp:Beukers2025}, where
\begin{equation}\label{eq:supp-block-hamiltonians}
\begin{aligned}
    \hat{\mathcal{H}}_{0}^{(0)} & = \gamma_c B_z \hat{I}_z,\\
    \hat{\mathcal{H}}_{0}^{(-1)} & = (\gamma_c B_z-A_\parallel)\hat{I}_z - A_\perp(\hat{I}_x\cos\theta_A+\hat{I}_y\sin\theta_A).
\end{aligned}
\end{equation}
Here and below, $\hat{I}_\alpha$ denotes the spin-$1/2$ nuclear operator acting within the conditional nuclear-spin block; the identity operator on the electron subspace is implicit after the block decomposition. These manifold-decomposed Hamiltonians define different nuclear-spin quantization axes. With $\omega_0=\gamma_cB_z$, the $m_s=0$ block is proportional to $\hat{I}_z$, so the corresponding quantization axis is aligned with the NV axis.

The $m_s=-1$ block can be written as
\begin{equation}\label{eq:supp-omega-one}
    \hat{\mathcal{H}}_{0}^{(-1)} = \omega_1\,\mathbf{n}_1\cdot\hat{\mathbf{I}}, \qquad \omega_1 = \sqrt{(\omega_0-A_\parallel)^2+A_\perp^2},
\end{equation}
where $\hat{\mathbf{I}}=\sum_{\alpha\in\{x,y,z\}}\hat{I}_\alpha\mathbf{e}_\alpha$, and the tilted quantization axis in the $m_s=-1$ manifold is described by the unit vector
\begin{equation*}
    \mathbf{n}_1 = \cos\beta_A\,\mathbf{e}_z-\sin\beta_A(\cos\theta_A\,\mathbf{e}_x+\sin\theta_A\,\mathbf{e}_y).
\end{equation*}
The azimuthal direction of the transverse component is set by $\theta_A$, with the sign convention inherited from the block Hamiltonian above.

For nonzero $\beta_A$, the eigenstates of $\hat{\mathcal{H}}_{0}^{(-1)}$ are spin states along $\mathbf{n}_1$, not the laboratory nuclear-spin states~\cite{supp:Taminiau2012,supp:Bradley2019}. An initial nuclear-spin state therefore precesses around the tilted $m_s=-1$ quantization axis. This manifold-dependent Larmor precession is the geometric origin of DD-induced conditional rotation and defines the hyperfine geometry shown in Fig.~1(b) of the main text.

\subsection{DD-induced conditional rotation}\label{sec:supp-dd-cr}
The DD-induced conditional rotation used in the main text follows the resonant control condition introduced for NV--nuclear-spin gates based on periodic MW $\pi$-pulses~\cite{supp:Taminiau2012,supp:Taminiau2014,supp:Bradley2019,supp:Dong2020}. For an ideal $(\tau-\pi-2\tau-\pi-\tau)$ block, where the MW $\pi$-pulses are treated as instantaneous, the inter-$\pi$-pulse interval, denoted by the superscript $(\mathrm{inst})$, is chosen as
\begin{equation}\label{eq:supp-ideal-dd-resonance}
    \tau_{\mathrm{wait}}^{(\mathrm{inst})} = \frac{(2k_{dd}-1)\pi}{\omega_0+\omega_1}, \qquad k_{dd}\in\mathbb{Z}^{+}.
\end{equation}
This condition synchronizes the nuclear precession accumulated in the $m_s=0$ and $m_s=-1$ manifolds so that the alternating quantization axes produce a conditional nuclear-spin rotation instead of an incoherent average over the two manifolds~\cite{supp:Taminiau2012,supp:Dong2020}.

\begin{figure}[t]\centering
    \includegraphics[width=1.0\textwidth]{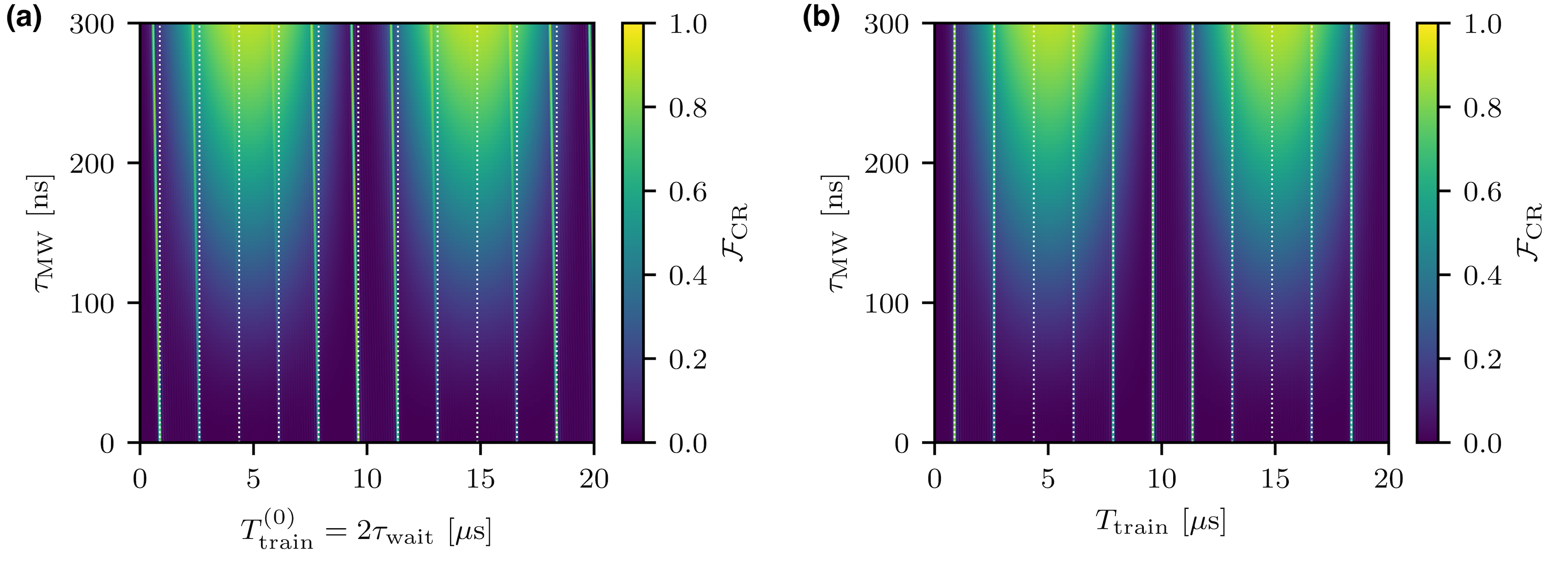}
    \caption{\label{fig:supp-mw-tau-compensation} Finite-MW-pulse timing compensation for the DD-induced CR condition with $N_{dd}=16$. The CR fidelity is calculated while varying the timing variable used to set the DD resonance condition. (a) Response plotted against the nominal train time $T_{\mathrm{train}}^{(0)}=2\tau_{\mathrm{wait}}$, which treats the MW $\pi$-pulse as instantaneous and therefore neglects the nuclear precession accumulated during the pulse. (b) The same response plotted against the corrected train time $T_{\mathrm{train}}=2\tau_{\mathrm{wait}}+\tau_{\mathrm{MW}}$. Including the finite MW-pulse duration restores the resonance positions, showing why $T_{\mathrm{train}}$ is used as the timing variable throughout this work.}
\end{figure}

In the experiments, the MW $\pi$-pulses have a finite duration $\tau_{\mathrm{MW}}$, and this finite pulse duration is included in the simulations throughout this study. During a MW pulse, the nuclear spin continues to precess under the static Hamiltonian. Therefore, relative to the instantaneous-pulse convention, a pulse centered between two free-evolution intervals contributes approximately half of its duration to each side, giving
\begin{equation}\label{eq:supp-wait-compensation}
    \tau_{\mathrm{wait}}^{(\mathrm{inst})} = \tau_{\mathrm{wait}} + \frac{\tau_{\mathrm{MW}}}{2}.
\end{equation}
Equivalently, defining the inter-$\pi$-pulse train time as
\begin{equation}\label{eq:supp-train-definition}
    T_{\mathrm{train}} = 2\tau_{\mathrm{wait}}+\tau_{\mathrm{MW}},
\end{equation}
automatically incorporates this finite-pulse compensation. Fig.~\ref{fig:supp-mw-tau-compensation} verifies the same point numerically using the CR-fidelity metric defined in \hyperref[sec:supp-cr-fidelity]{Sec.~II.B}. For short MW pulses the correction is small, but the residual timing error can accumulate as the number of $\pi$-pulses increases. Using $T_{\mathrm{train}}$ as the timing variable compensates this small dynamical shift by including the precession accumulated during the finite MW pulse.

Combining Eqs.~\eqref{eq:supp-ideal-dd-resonance}--\eqref{eq:supp-train-definition} gives
\begin{equation}\label{eq:supp-train-resonance}
    T_{\mathrm{train}} = 2\tau_{\mathrm{wait}}^{(\mathrm{inst})} = \frac{2(2k_{dd}-1)\pi}{\omega_0+\omega_1} = \frac{(2k_{dd}-1)\pi}{\bar{\omega}},
\end{equation}
where $\bar{\omega}=(\omega_0+\omega_1)/2$. This is the DD-induced conditional-rotation condition used in the main text.

The corresponding DD-induced rotation angle follows from the rotation angle accumulated per MW $\pi$-pulse in the standard DD-gate description~\cite{supp:Taminiau2012,supp:Bradley2019,supp:Dong2020,supp:VanOmmen2025}. With the finite-pulse correction absorbed into $T_{\mathrm{train}}$, this angle is
\begin{equation}\label{eq:supp-dd-angle-per-pulse}
    \cos\varphi_{dd} = \cos\beta_A\sin\frac{\omega_0T_{\mathrm{train}}}{2}\sin\frac{\omega_1T_{\mathrm{train}}}{2} - \cos\frac{\omega_0T_{\mathrm{train}}}{2}\cos\frac{\omega_1T_{\mathrm{train}}}{2}.
\end{equation}
Although the physical DD sequence uses an even number of MW $\pi$-pulses, the per-pulse angle $\varphi_{dd}$ is well defined independently of this parity. The total DD-induced rotation angle is therefore
\begin{equation}\label{eq:supp-total-dd-angle}
    \Phi_{dd}=N_{dd}\varphi_{dd}.
\end{equation}
In H-DDrf, $\Phi_{dd}$ is retained as a coherent offset, and the RF drive is used to supply the remaining operation. This value is the DD-induced angle used in Eq.~(1) of the main text.

\subsection{Geometric factor for RF driving}\label{sec:supp-geometric-factor}
The externally applied RF magnetic field is taken to oscillate along the $x$ axis of the laboratory nuclear-spin frame. To identify how the hyperfine geometry, specified by $\beta_A$ and $\theta_A$, affects RF driving in the $m_s=-1$ manifold, it is convenient to move from the laboratory nuclear-spin basis to the eigenbasis of $\hat{\mathcal{H}}_0^{(-1)}$. The corresponding eigenstates satisfy
\begin{equation*}
    \hat{\mathcal{H}}_0^{(-1)}|\psi_\pm\rangle = \pm\frac{\omega_1}{2}|\psi_\pm\rangle,
\end{equation*}
with
\begin{align*}
    |\psi_+\rangle & = \cos\frac{\beta_A}{2}|\!-\!1,\!\uparrow\rangle - e^{i\theta_A}\sin\frac{\beta_A}{2}|\!-\!1,\!\downarrow\rangle,\\
    |\psi_-\rangle & = e^{-i\theta_A}\sin\frac{\beta_A}{2}|\!-\!1,\!\uparrow\rangle + \cos\frac{\beta_A}{2}|\!-\!1,\!\downarrow\rangle.
\end{align*}
Here the arrows denote nuclear-spin states in the laboratory basis within the $m_s=-1$ electron manifold.

The basis transformation is generated by
\begin{equation}\label{eq:supp-full-rotation}
    \hat{R} = \exp\left[i\beta_A\hat{S}_z(\hat{I}_x\sin\theta_A-\hat{I}_y\cos\theta_A)\right],
\end{equation}
or equivalently, in electron-qubit block form, $\hat{R} = \hat{\mathbb{I}}_2\oplus\hat{\mathcal{R}}$ with $\hat{\mathcal{R}} = \exp[-i\beta_A(\hat{I}_x\sin\theta_A-\hat{I}_y\cos\theta_A)]$.
This gives $\hat{R}^{\dagger}\hat{H}_0\hat{R} = \omega_0\hat{I}_z\oplus\omega_1\hat{I}_z$. The same transformation maps the RF coupling operator as
\begin{equation*}
    \hat{R}^{\dagger}\hat{I}_x\hat{R} = \hat{I}_x\oplus\mathbf{C}\cdot\hat{\mathbf{I}},
\end{equation*}
where $\mathbf{C}=\sum_{\alpha\in\{x,y,z\}}C_\alpha\mathbf{e}_\alpha$ with
\begin{align*}
    C_x & = \sin^2\theta_A+\cos^2\theta_A\cos\beta_A, \\
    C_y & = \sin\theta_A\cos\theta_A(\cos\beta_A-1), \\
    C_z & = -\sin\beta_A\cos\theta_A.
\end{align*}
The longitudinal coefficient $C_z$ corresponds to a modulation of the Larmor frequency in the $m_s=-1$ eigenbasis. Within the secular approximation used here, this fast oscillating longitudinal contribution is averaged out, so that the effective RF coupling is evaluated with $C_z\to0$. The remaining transverse coefficients form an effective drive vector in the eigenbasis. Its magnitude determines the reduction of the RF Rabi rate, while the azimuthal angle added by the $C_y$ component determines the phase shift of the RF rotation axis relative to the applied laboratory-frame drive.

These two geometric effects are compactly described by
\begin{equation*}
    \mathcal{C}_\phi=C_x+iC_y,
\end{equation*}
so that $R_A = |\mathcal{C}_\phi|$ and $\delta\phi = \mathrm{arg}(\mathcal{C}_\phi)$, and the expanded form is Eq.~(3) in the main text. 

\subsection{RF pulse unitary operator}\label{sec:supp-rf-unitary}
In a standard rotating-frame treatment, an oscillating laboratory-frame drive becomes a simple rotation about a fixed transverse axis. For the present problem, we start from the electron-spin-rotating and nuclear-laboratory frame used for $\hat{H}_0$, transform to the nuclear eigenbasis defined by $\hat{R}$, move into the nuclear rotating frame, apply the RF-driven evolution, and finally transform back to the nuclear laboratory frame. For an RF pulse applied from $t_{\mathrm{on}}$ to $t_{\mathrm{off}}$, this sequence can be written as
\begin{equation}\label{eq:supp-rf-frame-sequence}
    \hat{U}_{\mathrm{RF}}(t_{\mathrm{on}},t_{\mathrm{off}};\phi_\mathrm{RF}) = \hat{R}\hat{U}_{\mathrm{ps}}(t_{\mathrm{off}})\hat{U}_{\mathrm{drive}}(t_{\mathrm{off}}-t_{\mathrm{on}},\phi_{\mathrm{RF}})\hat{U}_{\mathrm{ps}}^{\dagger}(t_{\mathrm{on}})\hat{R}^{\dagger},
\end{equation}
where $\hat{U}_{\mathrm{ps}}(t) = \exp(-i\omega_1\hat{I}_zt)$ accounts for the phase evolution in the nuclear eigenbasis, and $\hat{U}_{\mathrm{drive}}(\tau,\phi_{\mathrm{RF}}) = \exp(-i\hat{\tilde{H}}_{\mathrm{RF}}\tau)$ describes the RF-driven evolution in the eigenbasis rotating frame.

After the rotating-wave approximation, the RF Hamiltonian in this frame is $\hat{\tilde{H}}_{\mathrm{RF}} = \hat{\tilde{\mathcal{H}}}_{\mathrm{RF}}^{(0)}\oplus\hat{\tilde{\mathcal{H}}}_{\mathrm{RF}}^{(-1)}$, where
\begin{align}\label{eq:supp-rf-drive-hamiltonian}
    \hat{\tilde{\mathcal{H}}}_{\mathrm{RF}}^{(0)} & = -\delta_A\hat{I}_z + \frac{\Omega_{\mathrm{RF}}}{R_A}\left(\hat{I}_x\cos\phi_{\mathrm{RF}} + \hat{I}_y\sin\phi_{\mathrm{RF}}\right),\\
    \hat{\tilde{\mathcal{H}}}_{\mathrm{RF}}^{(-1)} & = \Omega_{\mathrm{RF}}\left(\hat{I}_x\cos(\phi_{\mathrm{RF}}+\delta\phi) + \hat{I}_y\sin(\phi_{\mathrm{RF}}+\delta\phi)\right).
\end{align}
The first term in the $m_s=0$ block appears because the same RF drive, resonant with the $m_s=-1$ nuclear transition, is detuned from the $m_s=0$ nuclear transition by $\delta_A=\omega_1-\omega_0$. The phase shift $\delta\phi$ and the dilation factor $R_A$ are the geometric corrections derived in \hyperref[sec:supp-geometric-factor]{Sec.~I.D}.

The RF drive amplitude is calibrated through the apparent RF Rabi frequency $\Omega_{\mathrm{RF}}$ experienced in the $m_s=-1$ manifold. In contrast, the nuclear quantization axis in the $m_s=0$ manifold is aligned with the laboratory $z$ axis, so no pulse-dilation factor is applied there. Therefore, an RF field chosen to produce $\Omega_{\mathrm{RF}}$ in the $m_s=-1$ manifold drives the $m_s=0$ manifold with the stronger amplitude $\Omega_{\mathrm{RF}}/R_A$, as written in Eq.~\eqref{eq:supp-rf-drive-hamiltonian}.

\subsection{RF pulse-duration condition}\label{sec:supp-rf-duration}
The RF drive is chosen to control the nuclear spin resonantly in the $m_s=-1$ manifold. In the $m_s=0$ manifold, the same drive is detuned by $\delta_A=\omega_1-\omega_0$ and has the actual Rabi frequency $\Omega_{\mathrm{RF}}/R_A$, as shown in Eq.~\eqref{eq:supp-rf-drive-hamiltonian}. The generalized Rabi frequency in the $m_s=0$ manifold is therefore
\begin{equation*}
    \Omega = \sqrt{\delta_A^2 + \frac{\Omega_{\mathrm{RF}}^2}{R_A^2}}.
\end{equation*}
The oscillation amplitude of the undesired detuned response is reduced by the factor $\Omega_{\mathrm{RF}}^2/(R_A^2\Omega^2)$, but the main requirement here is to close this off-resonant Rabi oscillation over an integer number of cycles. This gives
\begin{equation}\label{eq:supp-detuned-closure}
    \Omega\tau_{\mathrm{RF}}=2m\pi,
\end{equation}
where $m\in\mathbb{Z}^{+}$. The RF-assisted rotation angle required in the $m_s=-1$ manifold is $\Phi_{\mathrm{RF}}$, resulting in the RF Rabi frequency $\Omega_{\mathrm{RF}}=\Phi_{\mathrm{RF}}/(N_{dd}\tau_{\mathrm{RF}})$. Substituting this into Eq.~\eqref{eq:supp-detuned-closure} and solving for $\tau_{\mathrm{RF}}$ gives
\begin{equation}\label{eq:supp-rf-duration-general}
    \tau_{\mathrm{RF}} = \frac{\sqrt{4\pi^2N_{dd}^2R_A^2m^2 - \Phi_{\mathrm{RF}}^2}}{|\delta_A|N_{dd}R_A}.
\end{equation}
The fastest solution within this analytic pulse-timing condition is obtained by taking $m=1$, which gives Eq.~(4) in the main text. Further reductions of the gate duration may be possible through direct parameter optimization, but such numerical optimization is beyond the scope of the present work.

The RF pulses must also fit inside the free waiting window. Using the instantaneous-MW approximation for the inter-$\pi$-pulse waiting time,
\begin{equation*}
    \tau_{\mathrm{wait}} \simeq \frac{(2k_{dd}-1)\pi}{2\bar{\omega}},
\end{equation*}
and using the upper estimate $\tau_\mathrm{RF} \lesssim 2\pi/|\delta_A|$, the condition $\tau_{\mathrm{RF}}\leq\tau_{\mathrm{wait}}$ is satisfied when
\begin{equation*}
    \frac{2\pi}{|\delta_A|} \leq \frac{(2k_{dd}-1)\pi}{2\bar{\omega}}.
\end{equation*}
Solving this inequality gives
\begin{equation*}
    \frac{2\bar{\omega}}{|\delta_A|}+\frac{1}{2} \leq k_{dd}.
\end{equation*}
Thus the minimum train index used in the main text is
\begin{equation}\label{eq:supp-kdd-min}
    k_{dd,\mathrm{min}} = \left\lceil\frac{2\bar{\omega}}{|\delta_A|}+\frac{1}{2}\right\rceil.
\end{equation}

\section{Simulation and evaluation procedure}\label{sec:supp-simulation}

\subsection{Total DDrf sequence propagator}\label{sec:supp-propagator}

The numerical propagators are constructed from the elementary RF, MW, and free-evolution segments described below. The RF propagator derived in Eq.~\eqref{eq:supp-rf-frame-sequence} is denoted as $\hat{U}_{\mathrm{RF}}(t_{\mathrm{on}},t_{\mathrm{off}};\phi_{\mathrm{RF}})$ in the following. The finite MW $\pi$-pulse is described by
\begin{equation}\label{eq:supp-mw-pulse}
    \hat{U}_{\mathrm{MW}} = \exp\left[-i(\hat{H}_0+\hat{H}_{\mathrm{MW}})\tau_{\mathrm{MW}}\right],
\end{equation}
where $\hat{H}_{\mathrm{MW}}$ is the resonant MW control Hamiltonian introduced in \hyperref[sec:supp-frame]{Sec.~I.A}. In the simulations used in this work, $\tau_{\mathrm{MW}}=20~\mathrm{ns}$, corresponding to $\Omega_{\mathrm{MW}}/2\pi=25~\mathrm{MHz}$.

Between an RF pulse and an MW $\pi$-pulse, the system evolves freely under $\hat{H}_0$. Since $\tau_{\mathrm{RF}}\leq\tau_{\mathrm{wait}}$, the short free evolution segment is
\begin{equation}\label{eq:supp-short-free}
    \hat{U}_{\mathrm{free}} = \exp\left(-i\hat{H}_0(\tau_{\mathrm{wait}}-\tau_{\mathrm{RF}})\right).
\end{equation}
It is useful to combine the free evolution around each MW pulse into
\begin{equation}\label{eq:supp-u-pi}
    \hat{U}_{\pi} = \hat{U}_{\mathrm{free}}\hat{U}_{\mathrm{MW}}\hat{U}_{\mathrm{free}}.
\end{equation}
For comparison, the DD-only propagator is
\begin{equation*}
    \hat{U}_{\mathrm{DD}} = \left(e^{-i\hat{H}_0\tau_{\mathrm{wait}}}\hat{U}_{\mathrm{MW}}e^{-i\hat{H}_0\tau_{\mathrm{wait}}}\right)^{N_{dd}}.
\end{equation*}

The RF pulses in the DDrf sequence are applied with updated phases, and the propagator for the $n$-th pulse is defined as
\begin{equation}\label{eq:supp-rf-pulse-pieces}
    \hat{U}_{\mathrm{RF},n_{dd}} = \begin{cases} \hat{U}_{\mathrm{RF}}(0,\tau_{\mathrm{RF}};\phi_{\mathrm{RF},0}), & n_{dd}=0,\\ \hat{U}_{\mathrm{RF}}(n_{dd}T_{\mathrm{train}}-\tau_{\mathrm{RF}}, n_{dd}T_{\mathrm{train}}+\tau_{\mathrm{RF}}; \phi_{\mathrm{RF},n_{dd}}), & 1\leq n_{dd}\leq N_{dd}-1,\\ \hat{U}_{\mathrm{RF}}(N_{dd}T_{\mathrm{train}}-\tau_{\mathrm{RF}}, N_{dd}T_{\mathrm{train}}; \phi_{\mathrm{RF},N_{dd}}), & n_{dd}=N_{dd}.\end{cases}
\end{equation}
The RF phases follow the arithmetic update used for DDrf control,
\begin{equation}\label{eq:supp-rf-phase-update}
    \phi_{\mathrm{RF},n_{dd}} = n_{dd}\left(\pi-\frac{\delta_AT_{\mathrm{train}}}{2}\right) + \phi_0,
\end{equation}
where the H-DDrf implementation uses the geometrically matched initial phase $\phi_0=\theta_A-\delta\phi$. This phase-update rule is Eq.~(5) in the main text.

Including the MW $\pi$-pulses and the shortened free-evolution segments, the total DDrf-sequence propagator is
\begin{equation}\label{eq:supp-ddrf-propagator-expanded}
    \hat{U}_{\mathrm{DDrf}} = \hat{U}_{\mathrm{RF},N_{dd}}\hat{U}_{\pi}\hat{U}_{\mathrm{RF},N_{dd}-1}\hat{U}_{\pi}\hat{U}_{\mathrm{RF},N_{dd}-2}\cdots\hat{U}_{\mathrm{RF},1}\hat{U}_{\pi}\hat{U}_{\mathrm{RF},0}.
\end{equation}
Equivalently, using the time-ordering operator $\hat{\mathcal{T}}$ that places later sequence elements to the left, this may be written as
\begin{equation}\label{eq:supp-ddrf-propagator-product}
    \hat{U}_{\mathrm{DDrf}} = \hat{U}_{\pi}^{\dagger}\hat{\mathcal{T}}\left(\prod_{n_{dd}=0}^{N_{dd}}\hat{U}_{\pi}\hat{U}_{\mathrm{RF},n_{dd}}\right).
\end{equation}

\subsection{Conditional-operation contrast and CR-angle extraction}\label{sec:supp-cr-fidelity}
The full propagator after a DD or DDrf-type sequence can be projected onto the two electron-spin manifolds used as the NV qubit,
\begin{equation*}
    \hat{U} \simeq \hat{\mathcal{U}}_0 \oplus \hat{\mathcal{U}}_{-1},
\end{equation*}
where $\hat{\mathcal{U}}_0$ and $\hat{\mathcal{U}}_{-1}$ are the nuclear-spin operations conditioned on the NV state. The approximate equality becomes an exact block decomposition in the idealized limit where the MW $\pi$-pulses are treated as instantaneous. Since the target operation is conditional on the electron spin, we analyze the unitary blocks directly instead of propagating specific input states or evaluating reduced-state quantities. This reduces the computational cost and gives an immediate interpretation of whether the two electron manifolds generate distinct nuclear-spin operations. We therefore quantify the conditional part of the operation by comparing the two nuclear-spin blocks through their trace overlap, in analogy with the Hilbert-Schmidt overlap commonly used in gate-fidelity measures~\cite{supp:Nielsen2002,supp:Pedersen2007}. If $\hat{\mathcal{U}}_0$ and $\hat{\mathcal{U}}_{-1}$ are identical up to a global phase, the RF or DD sequence has produced an unconditional nuclear-spin operation. If their overlap is suppressed, the two electron manifolds generate distinguishable nuclear-spin operations and the sequence has a larger conditional component. We define the CR fidelity used in the main text as this conditional-operation contrast,
\begin{equation}\label{eq:supp-cr-fidelity}
    \mathcal{F}_\mathrm{CR} = 1-\frac{1}{4} \left|\mathrm{Tr}\!\left(\hat{\mathcal{U}}_{-1}^{\dagger}\hat{\mathcal{U}}_0\right)\right|^2 .
\end{equation}
This quantity is zero for an unconditional nuclear-spin operation and approaches unity for a maximally contrasting conditional operation.

The CR angle $\Phi_{\mathrm{CR}}$ plotted in the main text is extracted from the nuclear-spin rotation amplitudes generated separately in the two electron-spin manifolds. After removing irrelevant global phases, each conditional block can be treated as an $SU(2)$ rotation. For a rotation angle $\Phi_{\mathrm{CR},m_s}$ in the $m_s$ manifold, the trace relation gives
\begin{equation}\label{eq:supp-cr-angle-manifold}
    \cos\frac{\Phi_{\mathrm{CR},m_s}}{2}
    = \frac{1}{2}\mathrm{Tr}\!\left(\hat{\mathcal{U}}_{m_s}\right).
\end{equation}
Here, $m_s\in\{0,-1\}$. The CR angle reported in the figures is then the mean rotation amplitude generated by the two conditional nuclear-spin operations,
\begin{equation}\label{eq:supp-cr-angle}
    \Phi_{\mathrm{CR}} =
    \frac{\Phi_{\mathrm{CR},0}+\Phi_{\mathrm{CR},-1}}{2}.
\end{equation}
We refer to this averaged rotation amplitude as the CR angle when it is accompanied by a finite conditional-operation contrast $\mathcal{F}_\mathrm{CR}$, since the contrast verifies that the two electron-spin manifolds generate distinguishable nuclear-spin operations rather than the same unconditional rotation.

\subsection{Coherence-weight model}\label{sec:supp-coherence-weight}
The simulations in the main text focus on coherent gate construction rather than on a microscopic bath model. Instead of solving a Lindblad master equation or explicitly propagating the surrounding spin bath, we include decoherence through the experimentally established coherence envelope of DD-protected NV spins~\cite{supp:deLange2010,supp:Ryan2010,supp:Naydenov2011,supp:ShimDD2012,supp:BarGill2013}. This phenomenological treatment is sufficient for comparing gate protocols with different durations and DD-pulse numbers while keeping the coherent control mechanism transparent.

The coherence weight used in this work accounts for the loss of NV electron-spin coherence during the protected gate. Following the standard stretched-exponential description of DD-protected NV coherence, with the coherence time extended by the number of applied DD $\pi$-pulses, we define
\begin{equation}\label{eq:supp-coherence-weight}
    w_c = \exp\left[-\left(\frac{N_{dd}T_\mathrm{train}}{T_2^{(\mathrm{echo})}N_{dd}^{\chi_{dd}}}\right)^2\right].
\end{equation}
Here $T_2^{(\mathrm{echo})}$ is the NV coherence time measured by a Hahn-echo experiment, and $\chi_{dd}$ characterizes the DD-induced extension of the coherence time with the number of applied $\pi$-pulses. In the simulations, we use $T_2^{(\mathrm{echo})}=200~\mu\mathrm{s}$, representative of room-temperature NV centers, and $\chi_{dd}=0.54$, consistent with values used in NV DD coherence analyses~\cite{supp:deLange2010,supp:Ryan2010,supp:Naydenov2011,supp:ShimDD2012,supp:BarGill2013,supp:Beukers2025}. Nuclear-spin decoherence is not included separately, since the relevant $^{13}\mathrm{C}$ coherence times are much longer than the sub-millisecond gate durations considered here.

\subsection{RF-assisted rotation magnitude and sign selection}\label{sec:supp-rf-selection}
Eq.~(1) of the main text gives the RF-assisted rotation as $\Phi_{\mathrm{RF}}=\Phi_{\mathrm{goal}}-\Phi_{dd}$. In practice, the CR target is usually an odd multiple of $\pi/2$, and the useful RF correction is the smallest rotation that brings the DD-induced angle to the nearest such target. This avoids assigning an unnecessarily large RF rotation when the DD train has already brought the nuclear spin close to a successful CR operation.

We first define the wrapped angle
\begin{equation*}
    \mathrm{wrap}(\Phi) = \Phi - 2\pi\left\lfloor\frac{\Phi+\pi}{2\pi}\right\rfloor,
\end{equation*}
which maps $\Phi$ into the interval $(-\pi,\pi]$. Equivalently, in numerical implementation one may use $\mathrm{wrap}(\Phi)=\mathrm{mod}(\Phi+\pi,2\pi)-\pi$. Using the DD-induced angle $\Phi_{dd}=N_{dd}\varphi_{dd}$ from Eq.~\eqref{eq:supp-total-dd-angle}, the minimal RF-assisted correction to the nearest $\pm\pi/2$ target is chosen as
\begin{equation}\label{eq:supp-rf-angle-selection}
    \Phi_{\mathrm{RF}} = \mathrm{sgn}\!\left[\mathrm{wrap}(\Phi_{dd})\right]\frac{\pi}{2} - \mathrm{wrap}(\Phi_{dd}).
\end{equation}
This expression selects the RF rotation magnitude without explicitly searching over the integer in $\Phi_{\mathrm{goal}}=(2n-1)\pi/2$.

The sign of the RF drive also has to follow the DD-induced rotation direction. In the standard analysis of one $(\tau-\pi-2\tau-\pi-\tau)$ block~\cite{supp:Taminiau2012}, the conditional or unconditional rotation can be represented by an effective rotation axis. For a fixed axis, however, the actual DD-generated rotation can be parallel or antiparallel depending on the pulse timing and nuclear precession phases. The per-pulse DD angle $\varphi_{dd}$ in Eq.~\eqref{eq:supp-dd-angle-per-pulse} is extracted from an inverse cosine, so this directional information is not retained by the angle magnitude alone. The RF correction must therefore choose whether the RF phase follows the same axis or the opposite axis. Under the instantaneous MW $\pi$-pulse approximation, this is encoded by the direction factor
\begin{equation*}
    d_{\mathrm{RF}} = \mathrm{sgn}\!\left(-\frac{\omega_1T_{\mathrm{train}}}{2}\right).
\end{equation*}
As a result, the RF Rabi frequency used in the sequence is assigned as
\begin{equation}\label{eq:supp-rf-rabi-assignment}
    \Omega_{\mathrm{RF}} = \frac{d_{\mathrm{RF}}\Phi_{\mathrm{RF}}}{N_{dd}\tau_{\mathrm{RF}}}.
\end{equation}
Thus, $d_{\mathrm{RF}}$ determines the RF rotation direction and fixes the analytically assigned $\Omega_{\mathrm{RF}}$ after $\Phi_{\mathrm{RF}}$ has been selected. It does not modify the RF-pulse duration condition in Eq.~\eqref{eq:supp-rf-duration-general}, where $\Phi_{\mathrm{RF}}$ enters through its square.

\subsection{Detailed parameter settings for main-text figures}\label{sec:supp-main-fig-parameters}
We summarize the fixed and scanned quantities used for the main-text simulations. Unless otherwise stated, the pulse timings and RF phases are assigned by the analytic conditions derived in \hyperref[sec:supp-theory]{Sec.~I} and \hyperref[sec:supp-simulation]{Sec.~II}, rather than by numerical optimization over the full pulse sequence.

All simulations use $B_z=440.1~\mathrm{G}$, $\tau_{\mathrm{MW}}=20~\mathrm{ns}$, and $N_{dd}=16$, except for Figs.~2(c) and 2(d) of the main text where even values of $N_{dd}$ are scanned. The representative target spin is $(A_\parallel,A_\perp)/2\pi=(-200,50)~\mathrm{kHz}$ with $\theta_A=0$ and a target $\pi/2$ CR gate; when $\beta_A$ is scanned, $\delta_A/2\pi=220~\mathrm{kHz}$ is fixed and the hyperfine parameters are adjusted according to \hyperref[sec:supp-hyperfine-geometry]{Sec.~I.B}.

\begin{table}[h]
\caption{\label{tab:fig2-parameters}Train-time parameters used for Figs.~2(a) and 2(b) of the main text. H-DDrf uses the DD-induced CR condition, whereas DDrf uses the train time selected to suppress the DD-induced CR before RF driving.}
\begin{tabular}{@{}c|c|c@{}}
\hline
\hspace{0.5em}\textbf{Protocol}\hspace{0.5em} & \hspace{0.5em}\textbf{Parameter}\hspace{0.5em} & \hspace{0.5em}\textbf{Value}\hspace{0.5em} \\
\hline
\hline
\multirow[c]{2}{*}{H-DDrf}
  & $k_{dd}$ & $7$ \\
\cline{2-3}
  & $T_{\mathrm{train}}$ & $11.36250~\mu\mathrm{s}$ \\
\hline
\multirow[c]{2}{*}{DDrf}
  & $n_0$ & $3$ \\
\cline{2-3}
  & $T_{\mathrm{train}}$ & $12.73542~\mu\mathrm{s}$ \\
\hline
\end{tabular}
\end{table}

\begin{figure}[!t]\centering
    \includegraphics[width=0.85\textwidth]{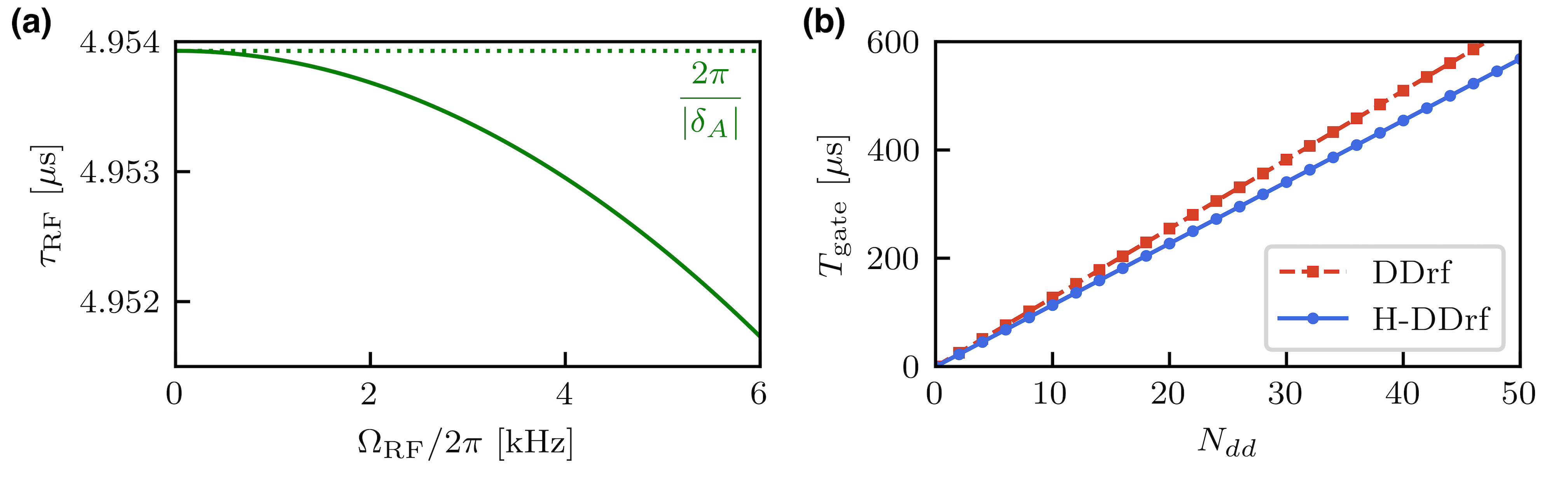}
    \caption{\label{fig:supp-fig2-timing} Timing parameters supporting Fig.~2 of the main text for the representative target spin $(A_\parallel,A_\perp)/2\pi=(-200,50)~\mathrm{kHz}$. (a) RF-pulse duration $\tau_{\mathrm{RF}}$ used for the RF-power scan in Figs.~2(a) and 2(b). The dotted line denotes the upper limit $2\pi/|\delta_A|$, while the solid curve is obtained by rearranging Eq.~\eqref{eq:supp-rf-duration-general} for the selected $\Omega_{\mathrm{RF}}$. (b) Gate duration $T_{\mathrm{gate}}=N_{dd}T_{\mathrm{train}}$ used for the DD-pulse-number scan in Figs.~2(c) and 2(d), with the DDrf and H-DDrf train times listed in Table~\ref{tab:fig2-parameters}. The vertical scale is chosen to match Fig.~2(d) of the main text.}
\end{figure}

For Figs.~2(a) and 2(b) of the main text, the fixed train times are listed in Table~\ref{tab:fig2-parameters}. For Figs.~2(c) and 2(d), the RF Rabi frequency is fixed at $\Omega_{\mathrm{RF}}/2\pi=1.50~\mathrm{kHz}$ while the number of DD $\pi$-pulses is varied. These values illustrate the timing used for the main-text comparison and should not be interpreted as a global minimum over all possible detuning and discrete timing conditions. Fig.~\ref{fig:supp-fig2-timing} shows the timing quantities that accompany the main-text scan; the sensitivity to RF-amplitude and RF-duration deviations is discussed separately in \hyperref[sec:supp-robustness]{Sec.~III.C}.

\begin{figure}[!t]\centering
    \includegraphics[width=1.0\textwidth]{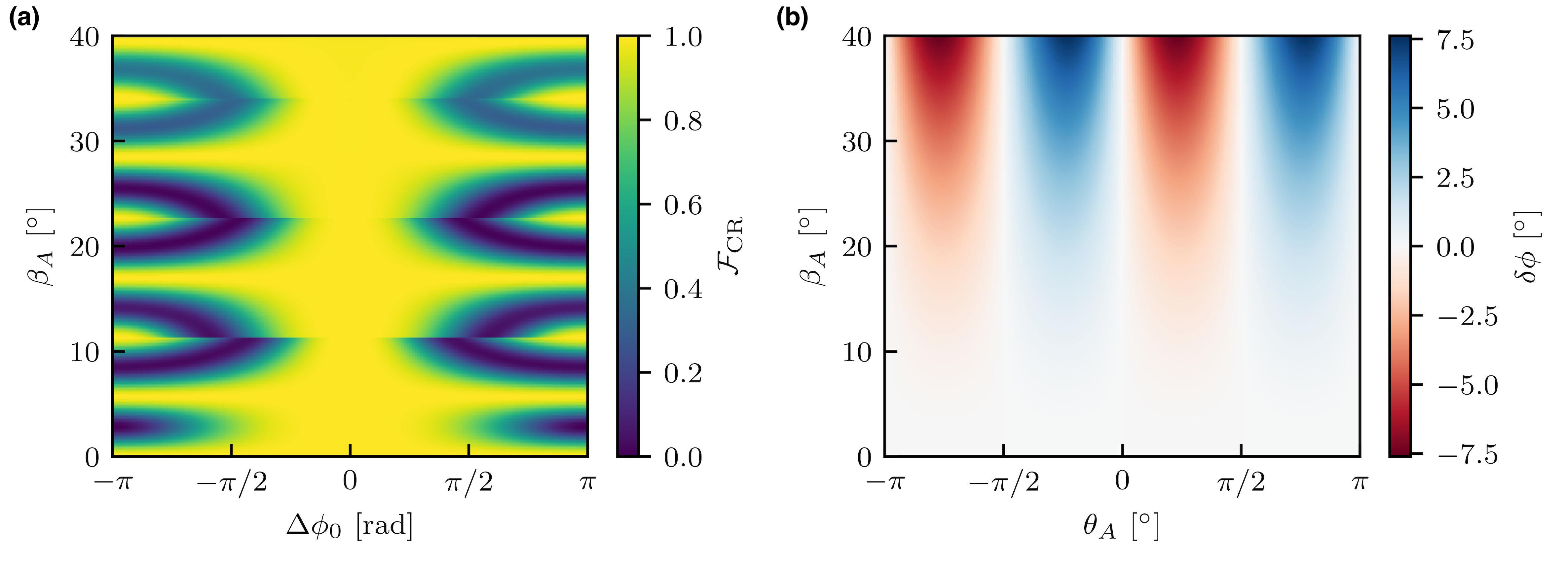}
    \caption{\label{fig:supp-fig3-parameters} Phase matching and geometric phase drag supporting Fig.~3(b) of the main text for fixed $\delta_A/2\pi=220~\mathrm{kHz}$. (a) $\mathcal{F}_{\mathrm{CR}}$ under H-DDrf as a function of the initial-phase displacement $\Delta\phi_0$ for a continuous range of $\beta_A$. The selected traces shown in the main text are taken from this continuous phase-matching map. For this detuning, the common train parameters are $k_{dd}=6$, $\tau_{\mathrm{wait}}=4.72218~\mu\mathrm{s}$, and $T_{\mathrm{train}}=9.46437~\mu\mathrm{s}$. (b) Dragged RF phase $\delta\phi=\mathrm{arg}(\mathcal{C}_\phi)$ as a function of the transverse-hyperfine azimuthal angle $\theta_A$ and quantization-axis misalignment $\beta_A$, again evaluated at $\delta_A/2\pi=220~\mathrm{kHz}$. This geometric phase determines the matched initial phase $\phi_0=\theta_A-\delta\phi$ used in panel (a).}
\end{figure}

Fig.~\ref{fig:supp-fig3-parameters}(a) shows the phase scan of Fig.~3(b) of the main text over a continuous range of $\beta_A$. The applied initial RF phase is written as $\phi_{\mathrm{applied}}=\phi_0+\Delta\phi$, where $\phi_0=\theta_A-\delta\phi$ is the geometrically matched value derived from $\mathcal{C}_\phi$. In regions where the DD-induced CR already nearly reaches the target operation, the required RF correction is small, $|\Omega_{\mathrm{RF}}|\simeq 0$, and the gate consequently becomes almost independent of the applied RF phase. Away from these DD-completed regions, increasing $\beta_A$ generally makes the successful phase window narrower, so the matched phase is needed for a deterministic parameter prescription across the hyperfine-geometry landscape. Fig.~\ref{fig:supp-fig3-parameters}(b) provides the corresponding geometric information by showing how the dragged phase $\delta\phi$ is generated by $\theta_A$ and $\beta_A$. The phase drag vanishes on the symmetry directions where the transverse hyperfine axis is aligned with or perpendicular to the RF-driving axis, while it becomes appreciable for generic azimuthal angles between these directions. This behavior follows from the imaginary component of $\mathcal{C}_\phi$, which is proportional to $\sin\theta_A\cos\theta_A$.

\subsection{Original-DDrf parameter-selection landscape}\label{sec:supp-bradley19-strategy}
For comparison with the original DDrf timing prescription, we evaluate the gate response over the discrete choices of the timing index $n_0$ and the number of DD $\pi$-pulses $N_{dd}$. This scan is not intended as an optimized benchmark of DDrf, but as a direct implementation of the baseline parameter-selection strategy described in Ref.~\cite{supp:Bradley2019}. The reference parameters are assigned within the small-transverse-coupling and instantaneous-MW-pulse approximation: the waiting and RF-pulse durations are set as $\tau_{\mathrm{wait}}=\tau_{\mathrm{RF}}=2\pi n_0/\omega_0$, giving $T_{\mathrm{train}}=4\pi n_0/\omega_0+\tau_{\mathrm{MW}}$ in our finite-MW-pulse notation, and the RF rotation is set without applying the geometric pulse-dilation factor $R_A$.

After these control parameters are assigned, the gate is evaluated using the same full Hamiltonian model as in the rest of this work, with the actual representative target spin $(A_\parallel,A_\perp)/2\pi=(-200,50)~\mathrm{kHz}$. Thus the parameter selection follows the original DDrf approximation, while the numerical propagation retains the finite transverse hyperfine interaction, manifold-dependent detuning, and finite MW $\pi$-pulse duration.

The purpose of this scan is to show how the discrete parameters of the baseline timing prescription enter the final gate response when the representative spin has a non-negligible transverse hyperfine component. Favorable parameter choices can still be found, but they appear as structured regions in the $(n_0,N_{dd})$ landscape and therefore require a parameter survey for each target spin. For this reason, the DDrf comparisons used elsewhere in this work employ the compensated DDrf construction, where the same geometric and timing corrections are applied before comparing with H-DDrf. This keeps the comparison focused on the hybrid use of the DD-induced CR offset, rather than on limitations of the original small-transverse-coupling approximation.

\begin{figure}\centering
    \includegraphics[width=0.57\textwidth]{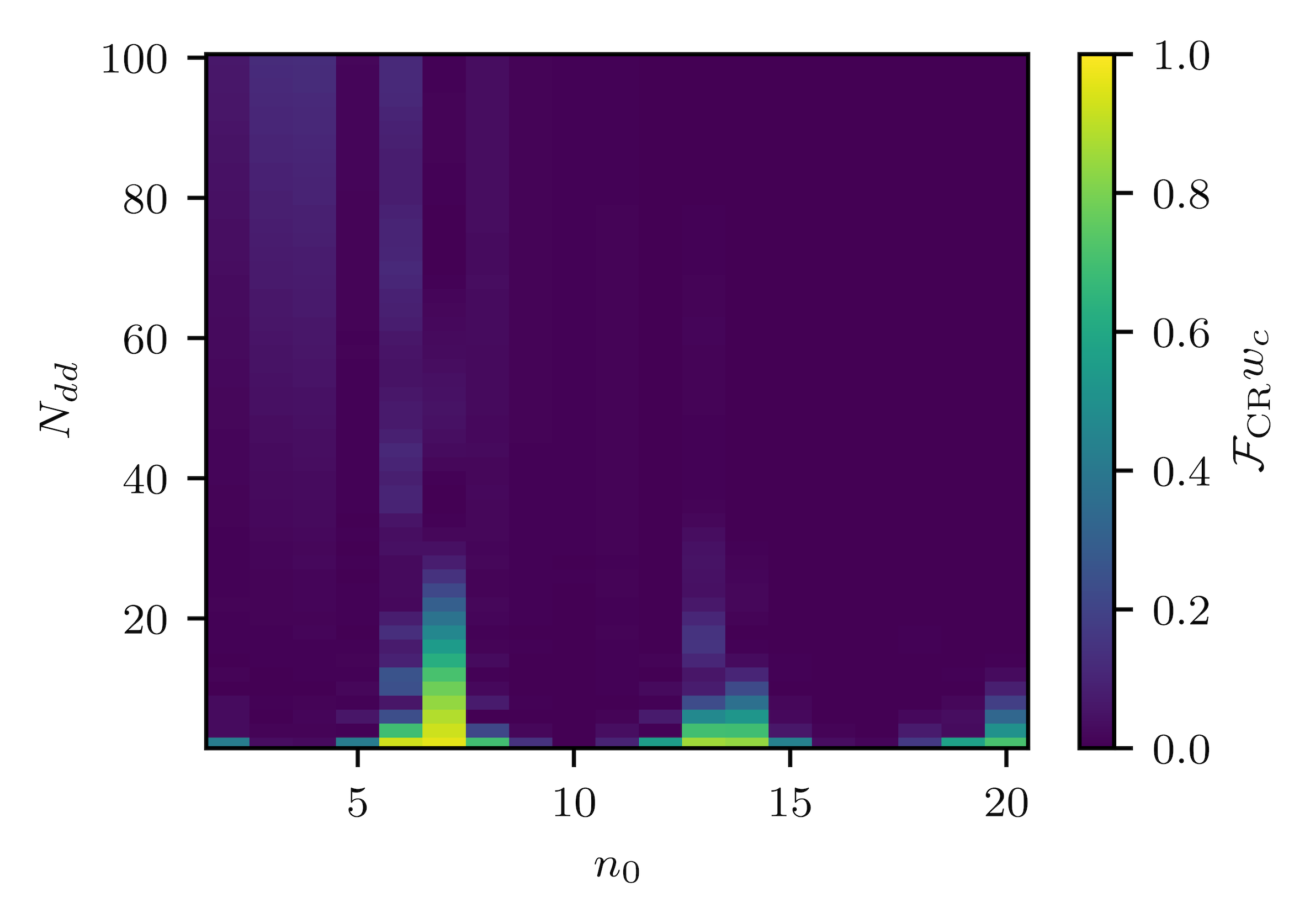}
    \caption{\label{fig:supp-bradley19-strategy} Parameter-selection landscape for the original DDrf timing prescription. The coherence-weighted CR fidelity $\mathcal{F}_{\mathrm{CR}}w_c$ is evaluated as a function of the DDrf timing index $n_0$ and the number of DD $\pi$-pulses $N_{dd}$ for $(A_\parallel,A_\perp)/2\pi=(-200,50)~\mathrm{kHz}$. Here, $n_0$ is the integer index that sets the original DDrf waiting and RF-pulse durations, $\tau_{\mathrm{wait}}=\tau_{\mathrm{RF}}=2\pi n_0/\omega_0$. The pulse parameters are assigned using the small-transverse-coupling prescription of Ref.~\cite{supp:Bradley2019}, while the resulting gate is propagated with the full Hamiltonian used throughout this work. Favorable operating points appear in selected regions of this discrete parameter space, so a parameter survey provides a natural way to identify suitable choices for a given target spin.}
\end{figure}

\section{Additional simulation results}\label{sec:supp-additional-results}

The following simulations supplement the main-text figures by separating three roles of the protocol. First, direct time-domain simulations show how the pulse sequence acts during the gate and how the nuclear spin evolves under the alternating MW and RF controls. Second, gate-level unitary simulations survey the deterministic parameter-selection rules over broader hyperfine-geometry and off-target-response ranges. Third, representative robustness checks test coherent deviations from the ideal RF controls. Together these results provide the bridge between the compact analytic rules in the main text and the numerical figures used to assess gate performance.

\subsection{Time-domain evolution simulations}\label{sec:supp-time-domain}
For the time-domain simulations, we perform direct numerical evolution in the partial rotating frame introduced in \hyperref[sec:supp-frame]{Sec.~I.A}. The MW control is applied as the resonant square-pulse Hamiltonian defined in \hyperref[sec:supp-propagator]{Sec.~II.A}, while the RF drive is kept explicitly time dependent,
\begin{equation*}
    \hat{H}_{\mathrm{RF},n_{dd}}(t) = 2\tilde{\Omega}_{\mathrm{RF}}\hat{I}_x \cos(\omega_1 t-\phi_{\mathrm{RF},n_{dd}}),
\end{equation*}
where $\tilde{\Omega}_{\mathrm{RF}}$ denotes the RF driving amplitude set by the applied RF magnetic field. The corresponding propagator is evaluated as the time-ordered exponential $\hat{U}(t_1,t_0)=\hat{\mathcal{T}}\exp\left[-i\int_{t_0}^{t_1}\hat{H}(t)\,dt\right]$, with $\hat{H}(t)$ including the static Hamiltonian, MW control, and RF drive terms active in each time segment. The explicitly time-dependent calculations use the QuTiP \texttt{sesolve} solver~\cite{supp:Johansson2012,supp:Johansson2013}, following the same operator definitions used in the gate-level simulations.

\begin{figure}\centering
    \includegraphics[width=1.0\textwidth]{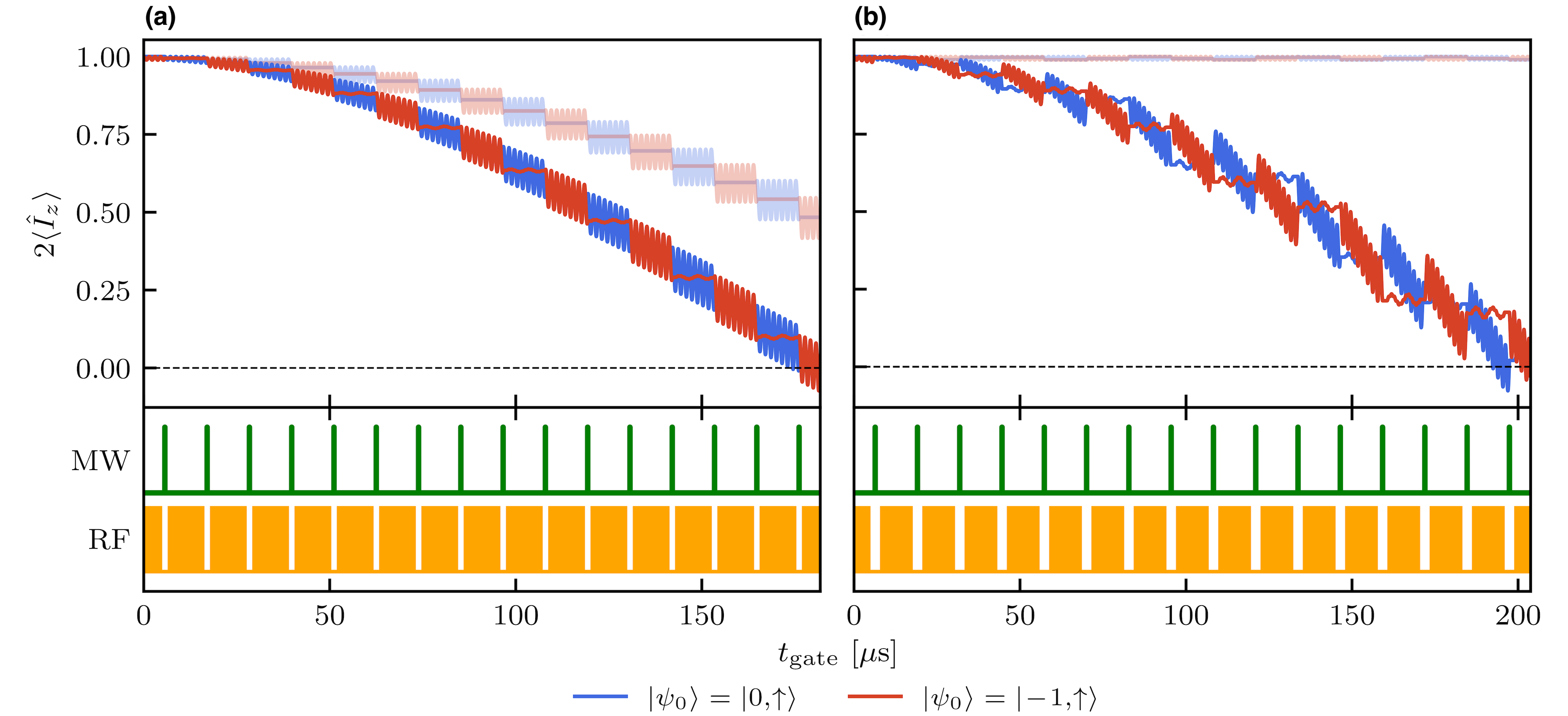}
    \caption{\label{fig:supp-time-evo-spin-z} Time-domain nuclear-spin evolution under the applied pulse sequence for the parameters in Table~\ref{tab:supp-time-evo-parameters}. The MW and RF pulse envelopes are shown together with the conditional nuclear spin-$z$ response obtained from direct time-dependent propagation. Panel (a) shows the H-DDrf sequence together with the DD-resonant RF-off evolution, while panel (b) shows the DDrf sequence together with the DD-suppressed RF-off evolution. The blue and red coherent curves correspond to the initially polarized nuclear state in the $m_s=0$ and $m_s=-1$ electron-spin manifolds, respectively, i.e., $|0,\!\uparrow\rangle$ and $|\!-\!1,\!\uparrow\rangle$. The transparent curves in both panels show the corresponding RF-off evolution with MW $\pi$-pulses only. The solid RF-assisted curves show the coherent conditional evolution generated after adding the RF pulses.}
\vspace{2.0em}
    \includegraphics[width=0.75\textwidth]{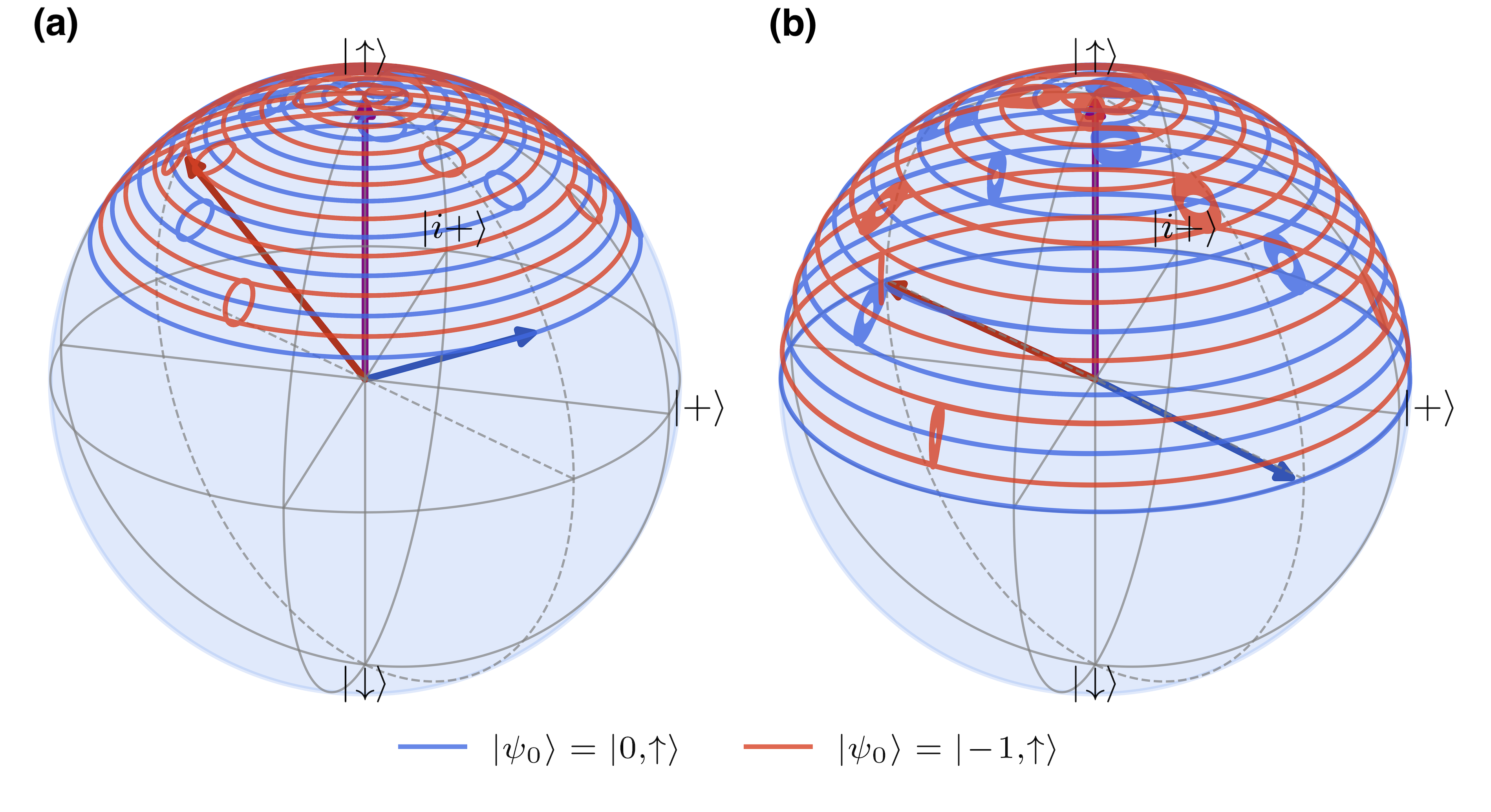}
    \caption{\label{fig:supp-time-evo-bloch} DD-only (a) and H-DDrf (b) time-domain evolutions visualized as conditional nuclear-spin trajectories on the Bloch sphere. The trajectories are shown in the nuclear rotating frame at $\omega_1$, so the rapid laboratory-frame Larmor precession is removed and the conditional gate trajectory is visible. In this frame, the $m_s=0$ trajectory retains the residual precession at the manifold detuning frequency $\delta_A=\omega_1-\omega_0$. The blue and red curves are the nuclear-spin-state trajectories initialized as $|0,\!\uparrow\rangle$ and $|\!-\!1,\!\uparrow\rangle$, respectively. Since the Bloch sphere represents only the nuclear-spin state, the initially polarized nuclear state is indistinguishable between the two electron-spin manifolds and is therefore shown as a single purple arrow. The blue and red trajectories start from the tip of this purple arrow, and their conditional final states are indicated by the corresponding final arrows. The dashed circular guides show that the final DD-only and H-DDrf rotations share the same azimuthal direction, which results from the initial geometric-phase matching of the RF drive.}
\end{figure}

The time-domain simulations are not used as the primary tool for broad parameter sweeps, because resolving every pulse segment is computationally less convenient than using the gate-level propagator. They are nevertheless useful as a direct check of the physical picture. Fig.~\ref{fig:supp-time-evo-spin-z} shows representative direct time-evolution simulations under the pulse Hamiltonians defined above. The plotted nuclear spin-$z$ component tracks the conditional response under H-DDrf and DDrf control together with the corresponding RF-off evolutions, while the aligned pulse sequence indicates the MW and RF segments used in the propagation. During the RF windows, the driven evolution inclines the nuclear spin continuously away from the stepwise DD-only trajectory. This smooth RF-assisted motion is added to the DD-generated offset in H-DDrf, whereas the DDrf trace starts from an RF-off trajectory with strongly suppressed DD contribution and therefore shows a more visibly segmented response in the spin-$z$ projection.

The same conditional evolution can be visualized as nuclear-spin trajectories on the Bloch sphere, as shown in Fig.~\ref{fig:supp-time-evo-bloch}. The trajectories are plotted in the nuclear rotating frame at $\omega_1$, so the rapid laboratory-frame Larmor precession is transformed away and the $m_s=-1$ manifold appears as a nearly stationary rotating-frame trajectory with only small residual circular motion. In the same frame, the $m_s=0$ trajectory retains the residual precession set by the manifold detuning $\delta_A=\omega_1-\omega_0$. The H-DDrf final arrow preserves the azimuthal direction of the DD-only rotation, showing that the RF-assisted part is added along the geometrically phase-matched rotation axis. The trajectory plot is therefore complementary to the spin-$z$ time trace: the time trace shows when the changes occur during the pulse sequence, whereas the Bloch-sphere representation shows how the nuclear trajectories conditioned on $m_s=0$ and $m_s=-1$ separate by the end of the operation.

\begin{table}[h]
\caption{\label{tab:supp-time-evo-parameters} Representative pulse parameters used for the direct time-domain simulations in Figs.~\ref{fig:supp-time-evo-spin-z} and \ref{fig:supp-time-evo-bloch}. Both simulations use $(A_\parallel,A_\perp)/2\pi=(-200,50)~\mathrm{kHz}$ and $N_{dd}=16$.}
\begin{tabular}{@{}c|c|c@{}}
\hline
\hspace{0.5em}\textbf{Parameter}\hspace{0.5em} & \hspace{0.5em}\textbf{H-DDrf}\hspace{0.5em} & \hspace{0.5em}\textbf{DDrf}\hspace{0.5em} \\
\hline
\hline
$T_{\mathrm{train}}$ & $11.36250~\mu\mathrm{s}$ & $12.73542~\mu\mathrm{s}$ \\
$\tau_{\mathrm{RF}}$ & $4.95387~\mu\mathrm{s}$ & $4.95332~\mu\mathrm{s}$ \\
$T_{\mathrm{gate}}$ & $181.80006~\mu\mathrm{s}$ & $203.76669~\mu\mathrm{s}$ \\
$\Omega_{\mathrm{RF}}/2\pi$ & $1.01473~\mathrm{kHz}$ & $3.15445~\mathrm{kHz}$ \\
$\mathcal{F}_{\mathrm{CR}}$ without RF & $0.76539$ & $0.00510$ \\
$\mathcal{F}_{\mathrm{CR}}$ with RF & $0.99999986$ & $0.99999908$ \\
\hline
\end{tabular}
\end{table}

\subsection{Gate-level unitary simulations}\label{sec:supp-gate-level}
The gate-level simulations use the total propagator in \hyperref[sec:supp-propagator]{Sec.~II.A} rather than resolving the full time-dependent trajectory. This makes it efficient to scan integer timing choices, hyperfine geometries, and off-target nuclear-spin parameters while using the same CR-fidelity and coherence-weight definitions as in the main text. In this section, each plotted point is generated by constructing the pulse parameters from the analytic rules and then evaluating the resulting propagator. The maps therefore test the parameter-selection procedure itself, not a locally optimized pulse sequence at each point.

\begin{figure}[t]\centering
    \includegraphics[width=0.5\textwidth]{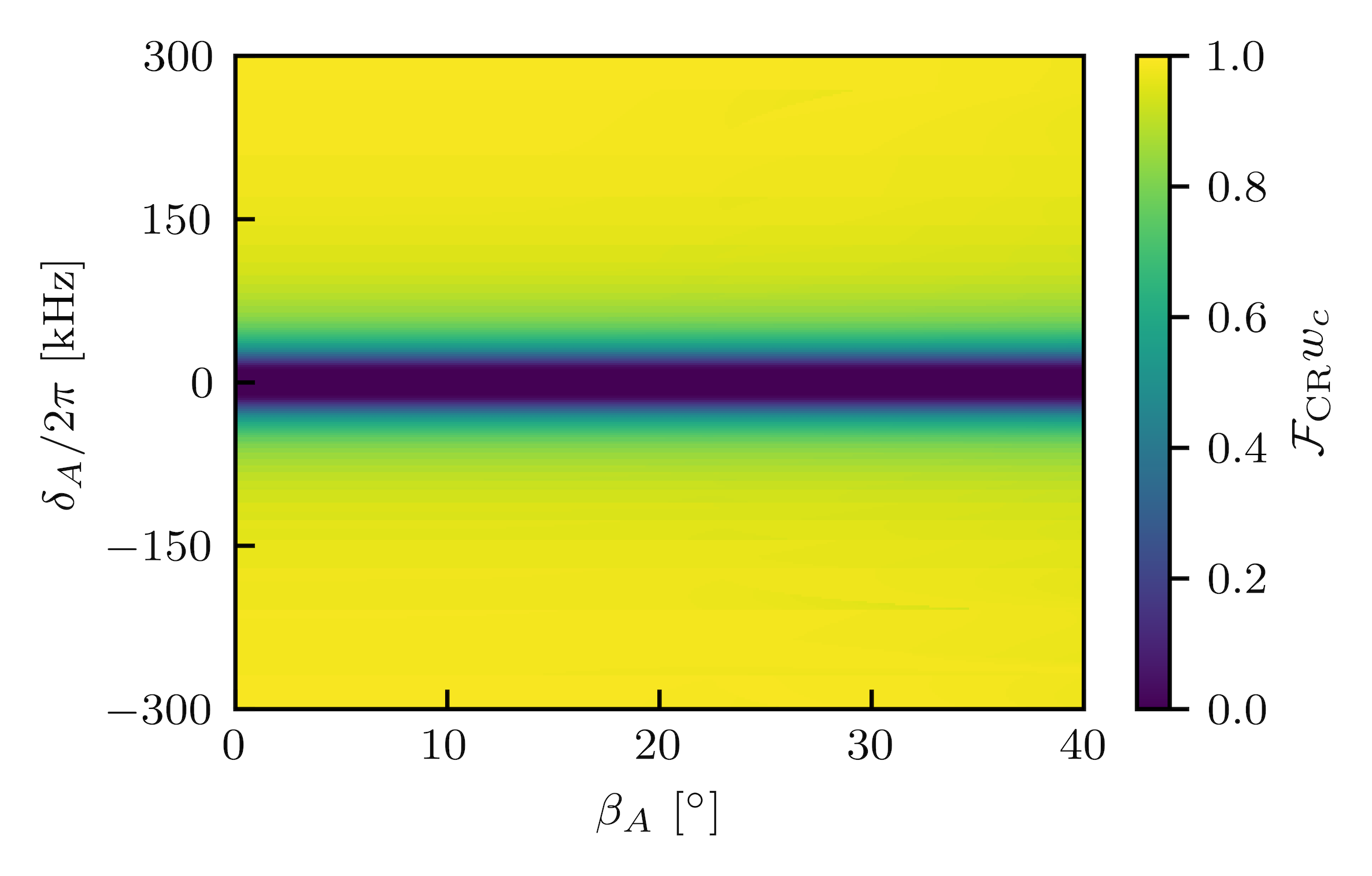}
    \caption{\label{fig:supp-hddrf-performance} Broad hyperfine-geometry applicability of H-DDrf. The coherence-weighted CR fidelity $\mathcal{F}_{\mathrm{CR}}w_c$ is plotted as a function of the manifold-dependent detuning $\delta_A$ and quantization-axis misalignment $\beta_A$. For each point, the H-DDrf control parameters are selected from the deterministic timing, RF-duration, and phase-matching rules derived in \hyperref[sec:supp-theory]{Sec.~I} and \hyperref[sec:supp-simulation]{Sec.~II}. The high-fidelity region extending to large $\beta_A$ shows that the protocol remains applicable beyond the small-misalignment regime. The low-fidelity band near $\delta_A=0$ reflects long gate durations and reduced selectivity when the two nuclear Larmor frequencies become nearly degenerate.}
\end{figure}
\begin{figure}[t]\centering
    \includegraphics[width=1.0\textwidth]{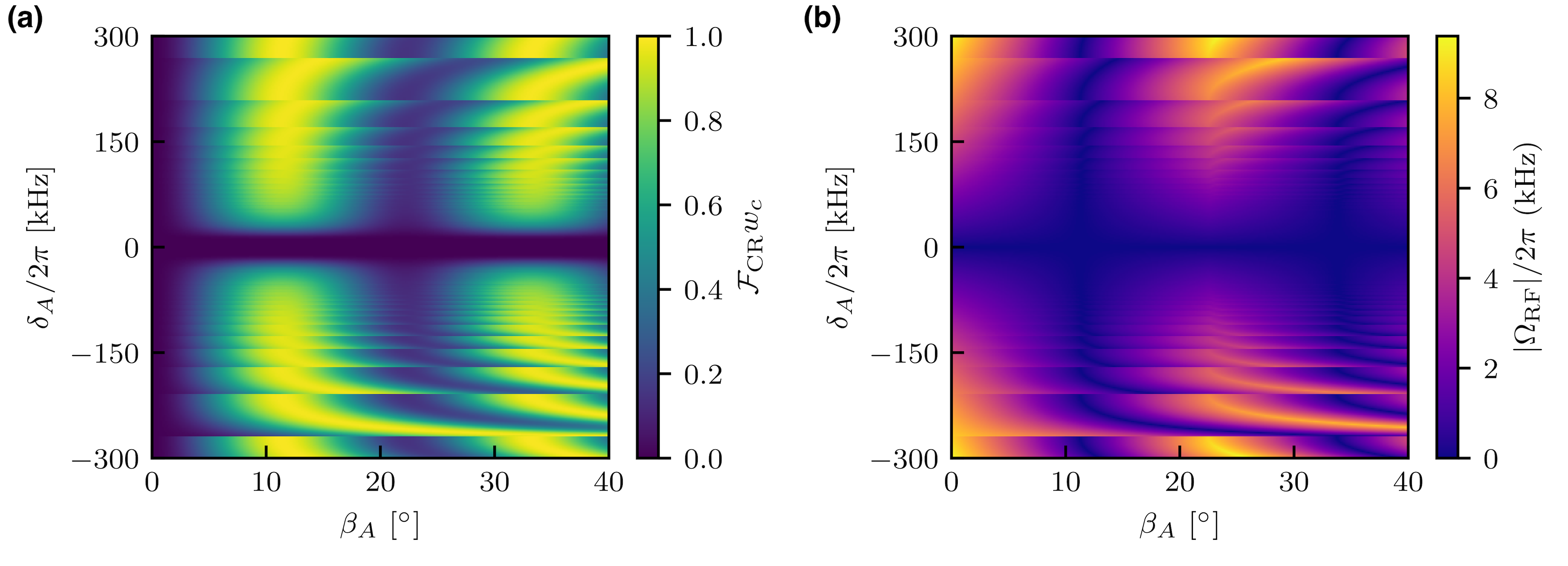}
    \caption{\label{fig:supp-no-rf-dd} Residual RF-assisted rotation required after DD-induced CR. The map shows the RF correction needed after the DD-induced angle $\Phi_{dd}$ is evaluated from Eq.~\eqref{eq:supp-total-dd-angle}. Regions where $\Phi_{dd}$ already lies close to the target CR angle require little or no RF-assisted correction according to Eq.~(1) of the main text. These regions illustrate the continuous connection between H-DDrf and DD-only control: the RF drive completes the gate only when the DD-induced offset does not already supply the desired operation. Two low-fidelity regions appear near $\beta_A=0$ and $\delta_A=0$: the former reflects the absence of a DD-induced CR when the quantization-axis misalignment is too small, while the latter reflects the coherence-weight loss caused by the long gate time required as the detuning approaches zero. The horizontal slices reflect stepwise changes of the selected $k_{dd,\mathrm{min}}$ condition in Eq.~\eqref{eq:supp-kdd-min} as a function of $\delta_A$.}
\end{figure}

\subsubsection{Broad hyperfine-geometry applicability of H-DDrf}\label{sec:supp-hddrf-performance-section}

Fig.~\ref{fig:supp-hddrf-performance} evaluates the deterministic H-DDrf prescription over a broad hyperfine-geometry landscape. For each point, the pulse train is selected from the minimum condition in Eq.~\eqref{eq:supp-kdd-min}, the RF duration and phase are assigned from the analytic rules in \hyperref[sec:supp-theory]{Sec.~I}, and the resulting gate is assessed using the coherence-weighted metric $\mathcal{F}_{\mathrm{CR}}w_c$. The map therefore tests whether the timing, phase-matching, and RF-assignment rules remain effective beyond the representative target spin used in the main-text examples.

The high-fidelity region extending to large $\beta_A$ shows that H-DDrf does not require the transverse hyperfine interaction to be treated as a small perturbation. This is the practical role of the geometric factors $R_A$ and $\delta\phi$: they incorporate the tilted nuclear quantization axis into the RF control prescription rather than requiring it to be suppressed. The main loss region appears near small $|\delta_A|$, where the two manifold-dependent nuclear transition frequencies become difficult to distinguish and the selected gate duration becomes long enough to be penalized by $w_c$.

\subsubsection{RF-unnecessary regions from DD-only completion}\label{sec:supp-no-rf-region}

In some hyperfine-geometry regions, the DD-induced CR angle already approaches the target operation, so Eq.~(1) of the main text requires only a small residual RF-assisted correction. In the limiting case where this residual angle vanishes, H-DDrf naturally reduces to a DD-only implementation. This is not a separate protocol, but a limiting case of the same hybrid rule: the RF drive is used only for the difference between the DD-induced angle and the desired CR angle.

This view is useful experimentally because it separates two kinds of control resources. The DD sequence provides spectral selection and a discrete conditional offset, while the RF drive provides continuous tunability around that offset. If a particular spin already lies close to a DD-only completion point, the required RF power can be reduced without changing the underlying DD resonance condition. The dark regions in Fig.~\ref{fig:supp-no-rf-dd} also clarify the limits of this reduction: small $\beta_A$ gives little DD-induced CR to use as an offset, and small $|\delta_A|$ requires long gates that are strongly penalized by the coherence weight.

\begin{figure}[t]\centering
    \includegraphics[width=1.0\textwidth]{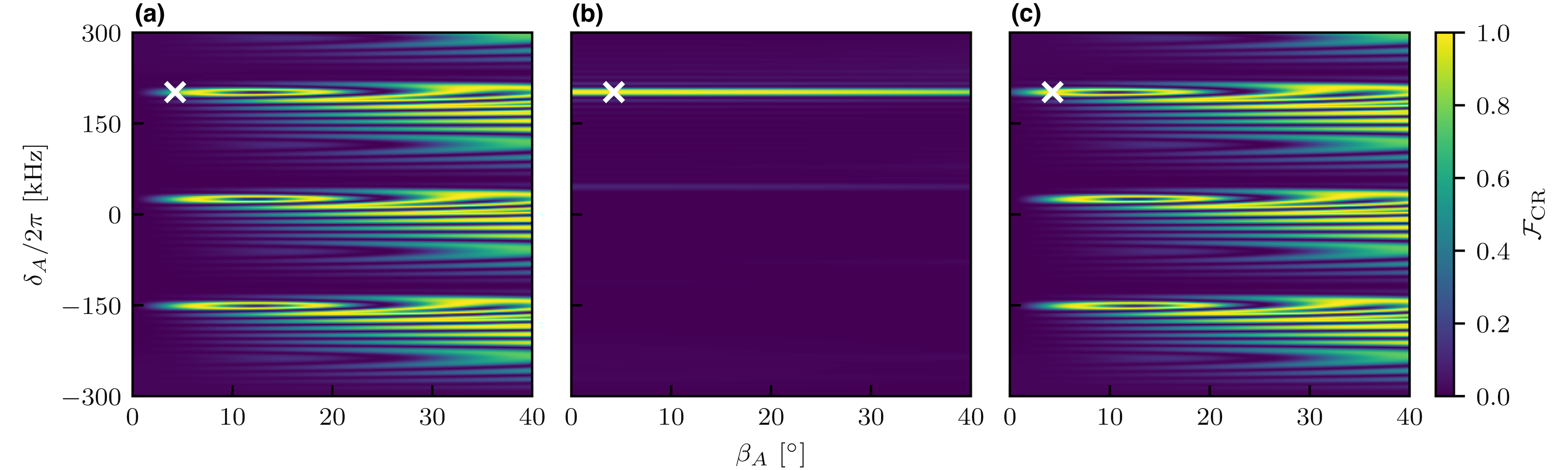}
    \caption{\label{fig:supp-harmonic-responses} Fidelity-harmonic response maps for (a) DD-only, (b) DDrf, and (c) H-DDrf control. The white cross marks the fixed target nuclear spin, $(A_\parallel,A_\perp)/2\pi=(-200,50)~\mathrm{kHz}$. For each point in the map, the control parameters are determined as if addressing a nuclear spin with that hyperfine geometry, while the propagated Hamiltonian is kept fixed to the target spin marked by the cross. The color scale therefore shows whether the target spin would respond when the sequence is tuned to another hyperfine condition. DD-only control exhibits multiple response bands associated with higher-order DD filter-function resonances. DDrf suppresses most DD-induced harmonic structure through RF selectivity, whereas H-DDrf inherits the DD-selected response and adds RF-assisted enhancement near the target spin. These maps illustrate that H-DDrf should be used together with the usual spectral-selection checks for nearby nuclear-spin registers.}
\end{figure}

\subsubsection{Selective enhancement and harmonic responses}\label{sec:supp-selective-harmonic}
The same gate-level simulations can be used to inspect harmonic responses away from the target spin. These responses are important because DD-based selectivity is not a single isolated resonance: higher-order filter-function resonances can also address other nuclear spins. Fig.~\ref{fig:supp-harmonic-responses} compares the fidelity-harmonic response for DD-only, DDrf, and H-DDrf control under the same target-spin setting. The rotation-harmonic response shows the same qualitative resonance structure, so we focus on the fidelity-harmonic response as the more direct gate-performance metric.

The comparison should be interpreted as a selectivity diagnostic rather than as a claim that all harmonic responses are removed. H-DDrf intentionally preserves the DD-selected conditional response and therefore can inherit some harmonic structure from the underlying DD sequence. Comparing the DD-only map in Fig.~\ref{fig:supp-harmonic-responses}(a) with the H-DDrf map in Fig.~\ref{fig:supp-harmonic-responses}(c) shows that the RF-assisted contribution acts as a spectral boost of the DD-selected bands: near the representative target detuning, the response pattern is shifted approximately parallel to the $\delta_A$ axis until the target spin reaches the desired CR completion. In the plotted H-DDrf response, the target spin becomes the absolute maximum, but the advantage of using the DD-induced evolution also carries the usual DD harmonic structure. In an experiment with several nearby nuclear spins, this map would guide the choice of the target spin and DD resonance condition before applying the RF-assisted completion.


\begin{figure}[t]\centering
    \includegraphics[width=0.86\textwidth]{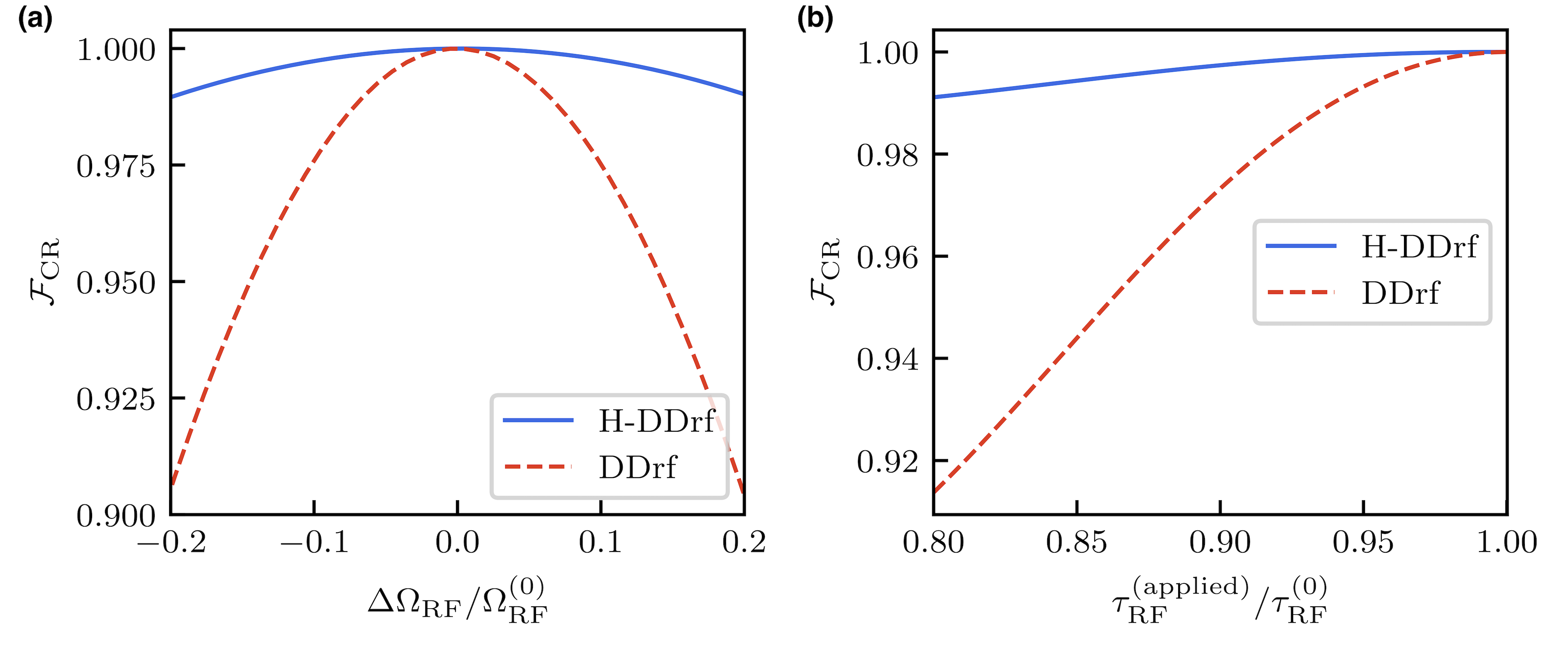}
    \caption{\label{fig:supp-omega-tau-robustness} Representative robustness to RF-control errors for the target spin $(A_\parallel,A_\perp)/2\pi=(-200,50)~\mathrm{kHz}$. (a) CR fidelity under a fractional RF-amplitude error $\Delta\Omega_{\mathrm{RF}}/\Omega_{\mathrm{RF}}$, with the pulse timings and RF phases fixed at their designed values. The central RF Rabi frequencies are $\Omega_{\mathrm{RF}}/2\pi=1.015~\mathrm{kHz}$ for H-DDrf and $3.154~\mathrm{kHz}$ for DDrf. (b) CR fidelity under an RF-duration deviation, with RF amplitude and phase sequence fixed while $\tau_{\mathrm{RF}}^{(\mathrm{applied})}/\tau_{\mathrm{RF}}^{(0)}$ is varied. Here $\tau_{\mathrm{RF}}^{(0)}=4.954~\mu\mathrm{s}$ is the analytic duration from Eq.~\eqref{eq:supp-rf-duration-general}. H-DDrf retains high CR fidelity because the RF drive supplies only the residual operation after the DD-induced CR offset.}
\end{figure}

\subsection{Representative robustness checks}\label{sec:supp-robustness}

The main text focuses on deterministic H-DDrf control parameters obtained from the timing and phase-matching conditions derived in \hyperref[sec:supp-geometric-factor]{Sec.~I.D}--\hyperref[sec:supp-rf-duration]{I.F} and \hyperref[sec:supp-rf-selection]{Sec.~II.D}. Here we outline representative coherent-error checks that test whether these conditions remain useful under small deviations from the ideal parameters. These checks are not intended to constitute a full experimental noise model; rather, they isolate the most relevant control-parameter sensitivities of the H-DDrf gate.

We separate calibration errors from parameter re-selection. In each robustness scan, the ideal pulse sequence is first constructed for the representative target spin. A single control parameter is then displaced while the other timing and phase parameters remain fixed. This procedure mimics a calibration error in the applied waveform, rather than a new analytic solution for a different target gate.

This subsection should be read as a continuation of the tunability, selectivity, and phase-matching results in Figs.~2 and 3 of the main text. The phase-matching robustness is already represented by Fig.~3(b) and Fig.~\ref{fig:supp-fig3-parameters}(a), where the applied initial phase is displaced from the geometrically matched value as
\begin{equation*}
    \phi_{\mathrm{applied}}=\phi_0+\Delta\phi,\qquad \Delta\phi\in[-\pi,\pi],
\end{equation*}
with $\phi_0=\theta_A-\delta\phi$. This scan directly probes whether the RF rotation axis remains aligned with the DD-induced CR axis. The tolerance to $\phi_{\mathrm{applied}}$ depends on the hyperfine geometry: in some regions the gate remains nearly phase-insensitive, whereas finite quantization-axis misalignment can narrow the allowed phase window. Even in the narrower cases shown in Fig.~3(b) of the main text and Fig.~\ref{fig:supp-fig3-parameters}(a), however, the response retains a finite tolerance around the matched phase rather than requiring an exact singular value. Similarly, Fig.~3(a) shows the frequency-selective response under an RF-frequency sweep. The detailed linewidth and off-resonant response depend on the applied RF power, so the frequency scan is used here as a representative selectivity test rather than as a universal power-independent robustness bound.

The remaining control-parameter checks are therefore RF-amplitude and RF-duration deviations around the designed operating point. For the RF-amplitude scan, the pulse duration and phase sequence are fixed at their designed values while only the applied RF amplitude is varied. RF-amplitude miscalibration is modelled as
\begin{equation*}
    \Omega_{\mathrm{RF}}\rightarrow (1+\epsilon_{\Omega})\Omega_{\mathrm{RF}},
\end{equation*}
with $\epsilon_{\Omega}$ scanned around zero. Since the RF drive supplies only the residual operation after the DD-induced CR offset, H-DDrf can reduce the absolute RF contribution required to reach the target gate in the representative regimes studied in the main text. The fractional sensitivity to $\epsilon_{\Omega}$, however, depends on the selected hyperfine parameters and target angle.

The RF-duration condition can be tested independently by fixing the RF amplitude and phase sequence while displacing only the applied RF-pulse duration,
\begin{equation*}
    \tau_{\mathrm{RF}}\rightarrow (1+\epsilon_{\tau})\tau_{\mathrm{RF}} .
\end{equation*}
Eq.~\eqref{eq:supp-rf-duration-general} is chosen to close the off-resonant Rabi response in the $m_s=0$ manifold. Deviations from this timing condition reopen the undesired response and reduce the conditional-operation contrast. The upper bound $\tau_{\mathrm{RF}}\lesssim 2\pi/|\delta_A|$ remains useful experimentally as a convenient initial guess before applying the exact duration condition. Fig.~\ref{fig:supp-omega-tau-robustness} summarizes these RF-amplitude and RF-duration checks for the representative target spin.

A full analysis of MW pulse-shape errors, composite-pulse compensation, and experimental noise spectra is beyond the scope of the present theoretical proposal.

\section{Diamond-lattice hyperfine-geometry survey}\label{sec:supp-diamond-survey}
The hyperfine-geometry survey in Fig.~4(a) of the main text is generated from a diamond-cubic lattice, which can be represented as an FCC Bravais lattice with a two-atom basis~\cite{supp:Kittel2005}. We place the NV center at the origin and choose the nearest nitrogen site along the crystallographic $[111]$ direction. The crystal-coordinate lattice vectors are first generated without rotation, and the coordinate frame is then rotated so that the NV axis, defined by this $[111]$ direction, becomes the laboratory $z$ axis used throughout the main text.

For each carbon site, we transform its position vector into this NV-axis frame and compute its radial distance and angular displacement from the NV center. We retain sites with $5~\text{\AA}\le r\le 7~\text{\AA}$, giving a representative shell of possible $^{13}\mathrm{C}$ registers around the defect. The lower bound excludes very close carbon sites, for which Fermi-contact contributions can be appreciable and a density-functional-theory treatment is generally required~\cite{supp:Gali2008}; these sites also produce MHz-scale hyperfine couplings outside the regime considered here. The upper bound avoids distant weakly coupled sites that behave predominantly as a spin bath and would concentrate the distribution near small hyperfine parameters~\cite{supp:Gali2008}. For each retained site, the hyperfine interaction is estimated from the electron-spin-secular part of the magnetic dipole-dipole interaction between the NV electron spin and the nuclear spin~\cite{supp:Doherty2013,supp:Dutt2007,supp:Maze2008}. We define the magnetic dipole-dipole interaction strength
\begin{equation*}
    J_0 = \frac{\mu_0\gamma_e\gamma_c\hbar}{4\pi},
\end{equation*}
which has units of angular frequency multiplied by distance cubed, with $J_0/2\pi\simeq19.879~\mathrm{kHz\,nm^3}$ in frequency units. For a carbon site at $\mathbf{r}_c=(x_c,y_c,z_c)$ in the NV-axis frame, with $\rho_c = \sqrt{x_c^2 + y_c^2}$ and $r_c = \sqrt{\rho_c^2 + z_c^2}$, the longitudinal and transverse hyperfine components are extracted as
\begin{equation}\label{eq:supp-dipolar-hyperfine}
    A_\parallel = J_0\frac{1-3(z_c/r_c)^2}{r_c^3},\qquad A_\perp = \frac{3J_0|z_c|\rho_c}{r_c^5}.
\end{equation}

Here $\theta_A=\mathrm{atan2}(y_c,x_c)=\mathrm{arg}(x_c+i y_c)$ is the azimuthal angle of the transverse hyperfine component in the $xy$ plane. The magnitude $A_\perp$ follows from the rotationally symmetric transverse component of the dipolar interaction, while $\theta_A$ is retained separately in the Hamiltonian in Eq.~\eqref{eq:supp-lab-hamiltonian} to specify the transverse direction. The resulting $(A_\parallel,A_\perp)$ values are then converted to the $(\delta_A,\beta_A)$ phase space shown in Fig.~4(a) of the main text.